\documentclass[twocolumn,english,amsart,showpacs,preprintnumbers,amsmath,amssymb,floatfix]{revtex4-1}

\usepackage{tikz,xcolor}
\usepackage[colorlinks = true,
linkcolor = blue,
urlcolor  = blue,
citecolor = blue,
anchorcolor = blue]{hyperref}
\definecolor{lime}{HTML}{A6CE39}
\DeclareRobustCommand{\orcidicon}{%
	\begin{tikzpicture}
		\draw[lime, fill=lime] (0,0)
		circle [radius=0.16]
		node[white] {{\fontfamily{qag}\selectfont \tiny ID}};
		\draw[white, fill=white] (-0.0625,0.095)
		circle [radius=0.007];
	\end{tikzpicture}
	\hspace{-2mm}
}
\foreach \x in {A, ..., Z}{%
	\expandafter\xdef\csname orcid\x\endcsname{\noexpand\href{https://orcid.org/\csname orcidauthor\x\endcsname}{\noexpand\orcidicon}}
}

\usepackage[T1]{fontenc}
\usepackage[latin9]{inputenc}
\usepackage{color}
\usepackage{array}
\usepackage{amstext}
\usepackage{graphicx}
\usepackage{esint}
\usepackage{rotating}
\usepackage[font=small,labelfont=bf]{caption}
\usepackage{xcolor}

\makeatletter

\@ifundefined{textcolor}{}
{%
 \definecolor{BLACK}{gray}{0}
 \definecolor{WHITE}{gray}{1}
 \definecolor{RED}{rgb}{1,0,0}
 \definecolor{GREEN}{rgb}{0,1,0}
 \definecolor{BLUE}{rgb}{0,0,1}
 \definecolor{CYAN}{cmyk}{1,0,0,0}
 \definecolor{MAGENTA}{cmyk}{0,1,0,0}
 \definecolor{YELLOW}{cmyk}{0,0,1,0}
 }

\@ifundefined{definecolor}
 {\usepackage{color}}{}
\@ifundefined{definecolor}
 {\usepackage{color}}{}
\makeatother
\usepackage{babel}

\begin{document}


\title {Effect of symmetry breaking of polarized light sea quarks on the nucleon and nuclear structure functions, and sum rules}

\author {Fatemeh Arbabifar$^{1}$\orcidA{}}
\email{F.Arbabifar@cfu.ac.ir}

\author {Shahin Atashbar Tehrani$^{2}$\orcidB{}}
\email{Atashbar@ipm.ir}

\author{Hamzeh Khanpour$^{3,4,2}$\orcidC{}}
\email{Hamzeh.Khanpour@cern.ch}

\affiliation {
$^{(1)}$Department of Physics, Nasibeh campus, Farhangian University, P.O.Box 14665-889, Tehran, Iran     \\
$^{(2)}$School of Particles and Accelerators, Institute for Research in Fundamental Sciences (IPM), P.O.Box 19395-5531, Tehran, Iran \\
$^{(3)}$AGH University of Science and Technology, Faculty of Physics and Applied Computer Science, Al. Mickiewicza 30, 30-055 Krakow, Poland \\
$^{(4)}$Department of Physics, University of Science and Technology of Mazandaran, P.O.Box 48518-78195, Behshahr, Iran
}

\date{\today}

%
\begin{abstract}\label{abstract}

In this study, we performed calculations and analyses of the structure functions 
of polarized nucleons and light nuclei, specifically $^3$He and $^3$H, using 
second-order Feynman diagrams. 
Our investigation focused on two main aspects: 
Firstly, we examined the symmetry properties of polarized light sea quarks. 
Secondly, we conducted a detailed investigation into the impacts of symmetry 
breaking on the structure functions of both nucleons and nuclei.
To achieve this, we utilized the existing polarized Parton Distribution 
Functions (polarized PDFs) available in the literature. These PDFs were used 
to calculate and compare the polarized structure functions $g_1$ and $g_2$ of the nuclei.
Additionally, we examined and analyzed the Bjorken and Efremov-Leader-Teryaev sum 
rules by utilizing the moments of the polarized structure functions.
The Lorentz color force components, namely $F_E^{y,n}$ and $F_B^{y,n}$, are 
determined using the twist-2, twist-3, and twist-4 matrix elements.
When symmetry breaking is applied, it is observed that they have 
similar magnitudes but opposite signs.
Our theoretical predictions for the polarized structure functions 
of nucleons and light nuclei, taking into account the symmetry breaking of 
light sea quarks, exhibit better agreement with experimental data.

\end{abstract}
%

\maketitle

%
%
\section{Introduction}\label{Introduction}

Over the past years, the Deep Inelastic Scattering (DIS) of leptons 
from nucleons has been a widely used experimental method for 
investigating the internal structure of nucleons.
Future lepton and hadron colliders, such as the Large Hadron-electron 
Collider (LHeC)~\cite{LHeC:2020van}, and the Future 
Circular Collider (FCC-he)~\cite{FCC:2018byv} will also play an important role 
in investigating the internal structure of nucleons~\cite{Hobbs:2020vme,Amoroso:2022eow}.
Experimental groups such as {\tt E142}, {\tt E143}, {\tt COMPASS}, {\tt HERMES}, 
and {\tt JLAB} have played a significant role in advancing our 
knowledge of nucleon structure by publishing a wealth of experimental results from collider experiments~\cite{E142:1996thl,E143:1998hbs,COMPASS:2015mhb,HERMES:2004zsh,JeffersonLabHallA:2003joy,JeffersonLabHallA:2016neg}. These experiments have focused on the DIS of polarized 
electrons by polarized nucleons and light nuclei, providing valuable 
insights into the nature of nucleon spin and its substructure.

The polarized structure functions of nucleons and nuclei provide useful 
information about the spin distribution of partons~\cite{Lin:2017snn,Ethier:2020way,Geesaman:2018ixo,Deur:2018roz,Zhou:2022wzm}.
In the simple picture of the $^3$He nuclei, all nucleons are
in the $S$ state wherein two protons with opposite spins exist,
hence, their spins in the asymmetry are completely canceled  and
the nuclei polarization is determined solely by the neutron spin.
Therefore, the use of $^3$He targets in DIS experiments of leptons from 
polarized targets is common and is considered as an alternative target to the neutron.
The same method applies in the case of $^3$H, where the neutrons are replaced with protons.
However, in more precise calculations and by considering other components of 
the three-particle wave, such as $S'$ and $D$ states, the spins of the protons 
are no longer canceled in the $^3$He structure function and must be taken into account.

In most published phenomenological models, the nucleon spin fractions carried by 
sea quarks are often assumed to be equal, and symmetry breaking is not taken into account.,  i.e. $\delta\bar{u}=\delta\bar{d}=\delta\bar{s}$~\cite{AtashbarTehrani:2007odq,Khorramian:2010qa,Taghavi-Shahri:2016idw,AtashbarTehrani:2013qea,Khanpour:2017cha,Salajegheh:2018hfs,Nematollahi:2021ynm,Mirjalili:2022cal,Sidorov:2006fi,Leader:2014uua,deFlorian:2005mw,Gluck:2000dy,Blumlein:2002qeu,deFlorian:2008mr,Hirai:2008aj,Blumlein:2010rn,Leader:2011tm,Leader:2010dx,Leader:2009tr,Nocera:2014uea}.
In some other phenomenological models, both flavor SU(2) and 
SU(3) symmetry breaking are taken into consideration, and hence,
the nucleon spin fraction carried by the light sea quarks considered to be unequal as $\delta\bar{u}\neq\delta\bar{d}\neq\delta\bar{s}$~\cite{Khanpour:2017fey,Leader:2010rb,deFlorian:2009vb,deFlorian:2014yva,Ball:2013lla,Arbabifar:2013tma,Khorramian:2020gkr,Lin:2017snn,Geesaman:2018ixo}.

In the present work, both of the mentioned phenomenological methods 
are applied, and the polarized structure functions of nucleons 
and nuclei are calculated using some selected polarized PDF 
models available in the literature, namely {\tt NAAMY21}~\cite{Nematollahi:2021ynm}, {\tt AKS14}~\cite{Arbabifar:2013tma}, {\tt DSSV09}~\cite{deFlorian:2009vb}, and 
{\tt BB10}~\cite{Blumlein:2010rn}.

In the most recent analysis, {\tt NAAMY21}, the polarized Deep Inelastic 
Scattering (DIS) data is utilized, and the polarized PDFs of protons, neutrons, and deuterons are 
calculated at the next-to-leading order (NLO) approximation, without 
taking into account symmetry breaking.
In other phenomenological model, {\tt AKS14}, the asymmetry data from 
inclusive and semi-inclusive polarized Deep Inelastic Scattering (SIDIS) are 
utilized, and both flavor SU(2) and SU(3) symmetry breaking are taken into consideration.
The {\tt QCD-PEGASUS} software package \cite{Vogt:2004ns} is employed 
in both analyses for the DGLAP (Dokshitzer-Gribov-Lipatov-Altarelli-Parisi) 
evolution, and a QCD fit is conducted using the experimental polarized data.
Both the {\tt NAAMY21} and {\tt AKS14} polarized PDFs are 
presented at the next-to-leading order (NLO) approximation in perturbative QCD.

In the present work, after calculating the moments of the polarized 
structure functions using the two mentioned polarized PDFs in the Mellin 
space, the DGLAP  evolution equations~\cite{Lampe:1998eu} are solved.
Then, using the Jacobi polynomials, the polarized structure
functions of nucleons are calculated in the Bjorken $x$ space.
Finally, the polarized structure functions of the light nuclei Helium-3 
($^3$He) and Tritium ($^3$H) are extracted in the second approximation of 
the Feynman diagram.
After applying the necessary corrections to the nuclei structure functions, 
we compare our results with the available experimental data for validation and comparison.
Having the polarized structure functions of nucleons and nuclei, along with 
their moments, one can calculate various important quantities such as the 
Bjorken sum rule, Efremov-Leader-Teryaev sum rule, and the Lorentz color force components.

The organization of this paper is as follows: In Section \ref{psf}, 
we present the theoretical background of this study, including 
the polarized PDFs, the polarized structure functions, and a comparison with 
the available experimental data.
In Section \ref{psfn}, we discuss the polarized structure functions 
of nuclei, the associated corrections, and compare the results with experimental data.
Sections \ref{bjorken} and \ref{elt} present our results on the sum rules, 
specifically the Bjorken sum rule and the Efremov-Leader-Teryaev sum rule. 
In Section \ref{lfc}, we perform the calculation of the Lorentz color force 
components. Finally, Section \ref{concl} provides a summary and conclusion of the work.

\section{Polarized structure functions of nucleons}\label{psf}

In this section, we will present the theoretical framework for the 
calculation of the polarized structure functions of nucleons in the Mellin space.
Following that, we will review the Jacobi polynomial method employed to 
transform the calculated structure functions from the Mellin space to the Bjorken $x$ space.
Additionally, we will provide a brief overview of previous studies on polarized PDFs, 
specifically the {\tt NAAMY21} and {\tt AKS14} models, which were utilized to 
compute the structure functions of nucleons and nuclei.

At the next-to-leading order (NLO) approximation, the polarized 
structure function of nucleons in Mellin space, 
denoted as ${\cal {M}} [x g_1, {\rm N}] (Q^2)$, can be expressed in 
terms of the polarized PDFs and the corresponding Wilson coefficient 
functions $\Delta C_i^N$. Therefore, the polarized structure function 
is given by the following expression:

\begin{eqnarray}
{\cal {M}} [x g_1^p, {\rm N}] (Q^2) & = &
\frac{1}{2}\sum\limits _{q=u,d,s}e_{q}^{2}\left\{
(1+\frac{\alpha_{s}}{2\pi}\Delta C_{q}^{N})\right.\nonumber \\
& \times & [\delta q(N,Q^{2})+\delta\bar{q}(N,Q^{2})]\nonumber \\
& + & \left.\frac{\alpha_{s}}{2\pi}\:2\Delta
C_{g}^{N}\delta g(N,Q^{2})\right\} \;,
\label{eq1}
\end{eqnarray}

\begin{eqnarray}
{\cal {M}} [x g_1^n, {\rm N}] (Q^2) & = &
{\cal {M}} [x g_1^p, {\rm N}] (Q^2)-\frac{1}{6}(1+
\frac{\alpha_{s}}{2\pi}\Delta C_{q}^{N})\nonumber\\
&\times&(\delta u(N,Q^{2})-\delta d(N,Q^{2}))\;, \label{eq2}	
\end{eqnarray}

\begin{eqnarray}
{\cal {M}} [x g_1^d, {\rm N}] (Q^2) & = &
\frac{1}{2}({\cal {M}} [x g_1^p, {\rm N}]+{\cal {M}} [x g_1^n, {\rm N}])\nonumber\\
&\times&\left(1-\frac{3}{2}\omega_D\right)\;, \label{eq3}	
\end{eqnarray}

for the proton ($p$), neutron ($n$) and deuteron ($d$), respectively.

In the equations above the $\omega_D=0.058$~\cite{Lacombe:1981eg,Buck:1979ff,Zuilhof:1980ae} and the
Wilson coefficient $\Delta C_{q}^{N}$ and $\Delta C_{g}^{N}$ can be
found in~\cite{Lampe:1998eu}, and written as follows,

\begin{eqnarray}
\Delta C_{q}^{N} & = & \frac{4}{3}\:\left\{ -S_{2}(N)+(S_{1}(N))^{2}+
\left(\frac{3}{2}-\frac{1}{N(N+1)}\right)\right.\nonumber\\
& \times & \left.S_{1}(N)+\frac{1}{N^{2}}+\frac{1}{2N}+
\frac{1}{N+1}-\frac{9}{2}\right\}\label{eq4} ,
\end{eqnarray}

\begin{eqnarray}
\Delta C_{g}^{N} & = & \frac{1}{2}\left[-\frac{N-1}{N(N+1)}(S_{1}(N)+1)-
\frac{1}{N^{2}}+\frac{2}{N(N+1)}\right] ,\nonumber\\\label{eq5}
\end{eqnarray}

with $S{}_{1}(n)=\sum_{j=1}^{n}\frac{1}{j}=\psi(n+1)+\gamma_{E}$,
$S{}_{2}(n)=\sum_{j=1}^{n}\frac{1}{j^{2}}=(\frac{\pi^{2}}{6})-\psi'(n+1)$,
$\psi(n)=\Gamma'(n)/\Gamma(n)$, $\gamma_{E}=0.577216$  and $\psi'(n)=d^{2}\ln\Gamma(n)/dn^{2}$.

It is important to note that the structure function of neutrons, 
defined through Eq.~\ref{eq2}, takes into account symmetry breaking, 
meaning that the contribution of sea quarks is considered.
However, if the symmetry of light sea quarks is neglected, 
the second term subtracted from the proton structure function 
can be replaced by $\delta u_v(N,Q^{2}) - \delta d_v(N,Q^{2})$.

After determining the moments of the structure functions, they can 
be transformed to the Bjorken $x$ space using the 
Jacobi polynomials \cite{Khorramian:2010qa}. The transformation is 
given by the following expression:

\begin{equation}\label{eq:xg1QCD}
	x \, g_{1}(x, Q^2) = x^{\beta} (1 - x)^{\alpha}\ \,
	\sum_{n = 0}^{\rm N_{ \rm max}} a_n(Q^2)
	\, \Theta_n^{\alpha, \beta}(x) \,,
\end{equation}

where the maximum order of expansion is denoted by $\rm N_{\rm max}$, and 
the index $n$ refers to the order of expansion. 
Two free parameters $\alpha$ and $\beta$ are chosen to make the fastest
convergence for the series in Eq.~\ref{eq:xg1QCD}.
The $a_{n}(Q^{2})$ as Jocbi moment discloses the dependence of
polarized structure functions on Q$^2$.
By using the Jacobi polynomials, one can factor out from the structure 
function a weight function $w^{\alpha, \beta}(x) \equiv  x^{\beta} (1 - x)^{\alpha}$. 
Hence, the Jacobi polynomials $\Theta_n^{\alpha, \beta}(x)$ can be written as the following expansion:

\begin{equation}
\label{eq:Jacobi}
\Theta_n^{\alpha, \beta}(x) = \sum_{j = 0}^{n}
\, c_j^{(n)}(\alpha, \beta) \, x^j \,.
\end{equation}

It can be shown that the expansion coefficient $c_j^{(n)}(\alpha, \beta)$ is given by a
combination of Gamma $\Gamma$  function in terms of $n$, $\alpha$ and $\beta$ parameters~\cite{Parisi:1978jv}.

The following orthogonality relation is satisfied by the 
Jacobi polynomials with the weight function $w^{\alpha, \beta}(x)$,

\begin{equation}\label{eq:orthogonality}
\int_0^1 dx \, x^{\beta} (1 - x)^{\alpha} \,
\Theta_n^{\alpha, \beta}(x)
\, \Theta_{l}^{\alpha, \beta}(x)
\, = \, \delta_{n, l} \,.
\end{equation}

Based on the orthogonality condition, the Jacobi moments $a_n(Q^2)$ can be
obtained as~\cite{Khorramian:2010qa,Kataev:2001kk,Kataev:2005ci}:

\begin{eqnarray}
\label{eq:aMoment}
a_n(Q^2) & = & \int_0^1 dx \, x g_1(x,Q^2) \,
\Theta_n^{\alpha, \beta}(x)   \nonumber \\
& = & \sum_{j = 0}^n \, c_j^{(n)}(\alpha, \beta)
\, {\cal M} [xg_{1}, \, j + 2](Q^2)  \;,
\end{eqnarray}

where the ${\cal M} [xg_{1}, \, j + 2](Q^2)$ is the Mellin transform
of the $xg_1 (x, Q^2)$ from Eqs.~\ref{eq1}, \ref{eq2} and \ref{eq3},  
which is given by, 
\begin{eqnarray}
\label{eq:Mellin-transform}
{\cal M} [xg_{1}, \, N] (Q^2) =  \int_0^1 dx \, x^{N-2} \, x g_{1} (x, Q^2)    \;,
\end{eqnarray}

Now by substituting Eq.~\ref{eq:aMoment} into Eq.~\ref{eq:xg1QCD}
the polarized structure
function $x g_1(x, Q^2)$, based on Jacobi polynomial expansion method,
can be constructed.
Therefore the following
expression for $x g_1(x, Q^2)$ could be obtained,

\begin{eqnarray}\label{eq6}
	x g_1(x, Q^2) & = & x^{\beta}(1 - x)^{\alpha} \,
	\sum_{n = 0}^{\rm N_{max}} \, \Theta_n^{\alpha, \beta}(x) \nonumber \\
	& \times & \sum_{j = 0}^n \, c_j^{(n)}{(\alpha, \beta)}
	\, {\cal M}[x g_1, j + 2](Q^2) \,.
\end{eqnarray}

where $N_{max}$ is considered to be 9 , and $\alpha$ and  $\beta$ are fixed to 3 and 0.5, respectively~\cite{Khorramian:2010qa,Khanpour:2017cha,Salajegheh:2018hfs,Khanpour:2017fey}.

\begin{figure}[!htb]
\includegraphics[clip,width=0.4\textwidth]{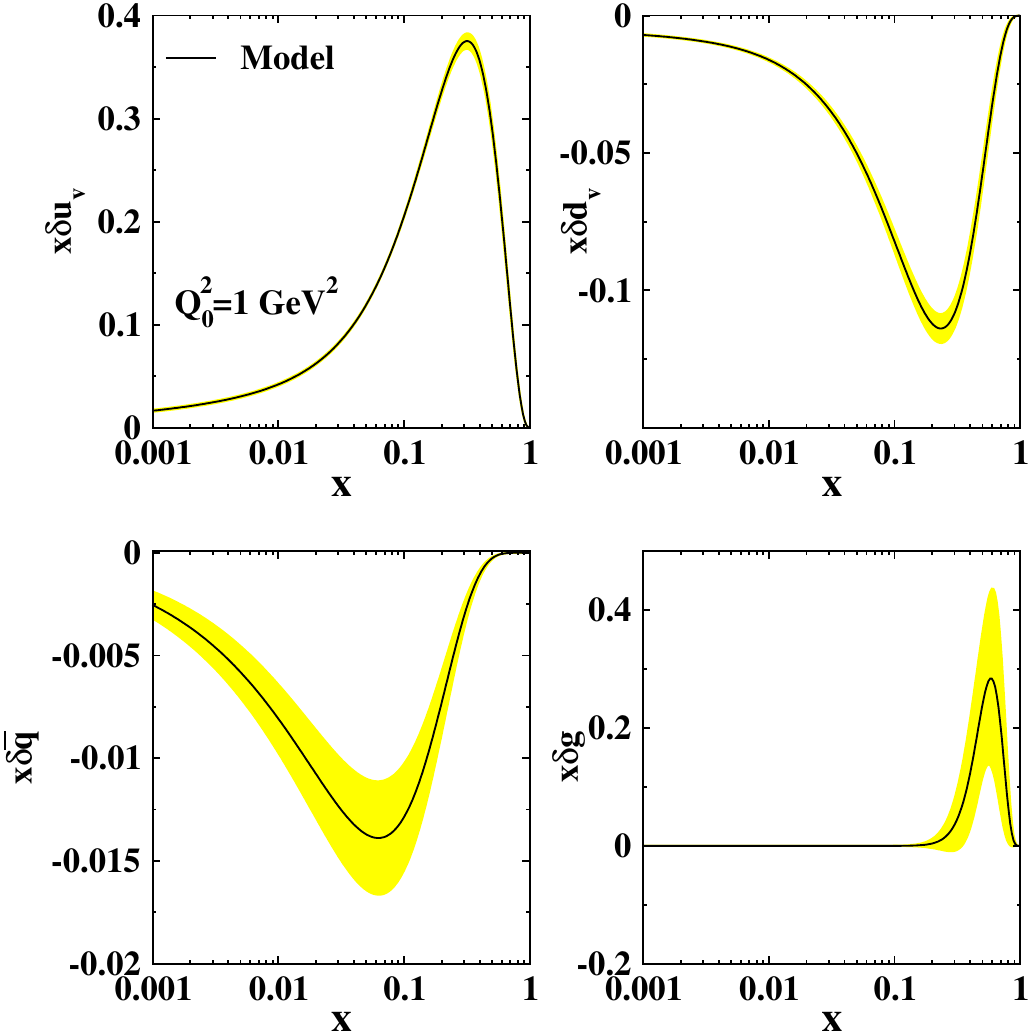}
\begin{center}
\caption{{\small The polarized PDFs from {\tt NAAMY21}~\cite{Nematollahi:2021ynm}
model at $Q_0^2 = 1  \, \textrm {GeV}^2$ in NLO approximation,
considering $\delta \bar{q}=\delta \bar{u}=\delta \bar{d}=\delta \bar{s}$.  \label{fig:fig1}}}
\end{center}
\end{figure}

\begin{figure}[!htb]
	\includegraphics[clip,width=0.4\textwidth]{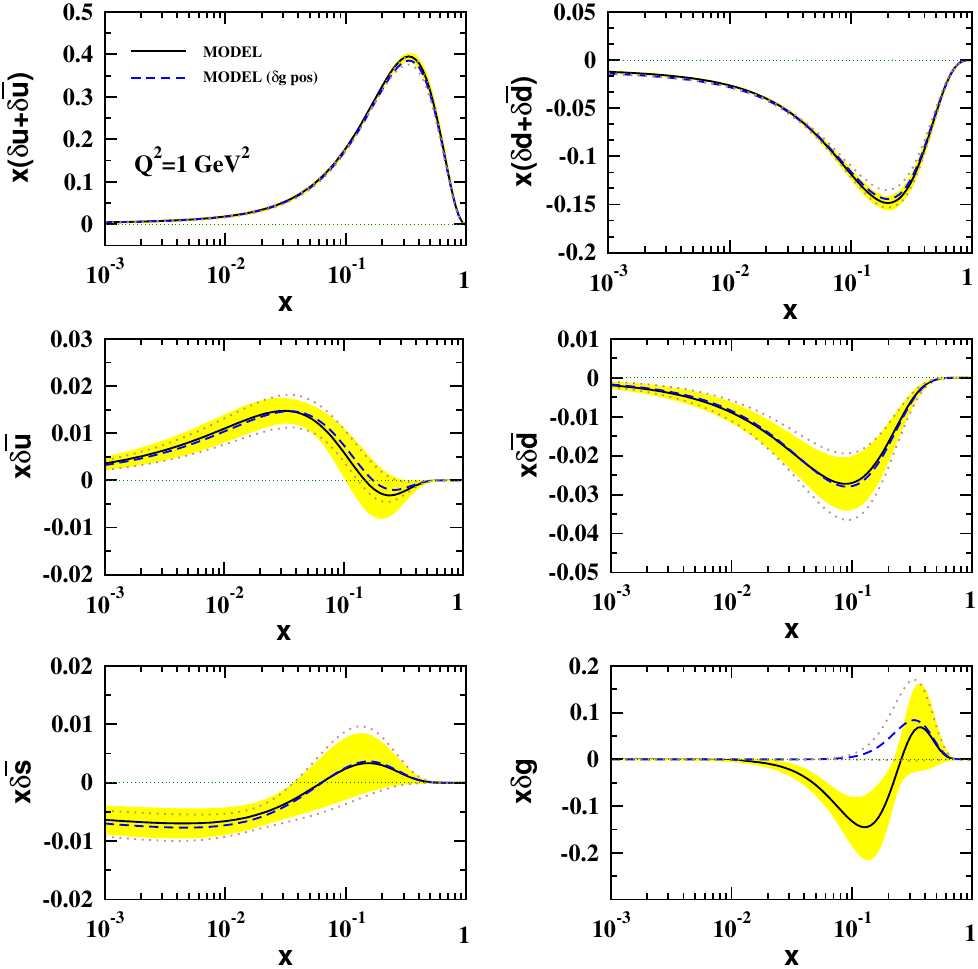}
	\begin{center}
\caption{{\small The polarized PDFs from {\tt AKS14}~\cite{Arbabifar:2013tma} model
at $Q_0^2=1  \,  \textrm{GeV}^2$ at NLO approximation,
considering $\delta \bar{u} \neq \delta \bar{d} \neq \delta \bar{s}$.
Solid line indicates the {\tt AKS14} model in sign changing gluon scenario,
dashed line denotes the {\tt AKS14} model in positive gluon scenario. \label{fig:fig2}}}
\end{center}
\end{figure}

The polarized PDFs of the {\tt NAAMY21} and  {\tt AKS14} are 
depicted in Figures \ref{fig:fig1} and \ref{fig:fig2}, respectively.

\begin{figure}[!htb]
\includegraphics[clip,width=0.45\textwidth]{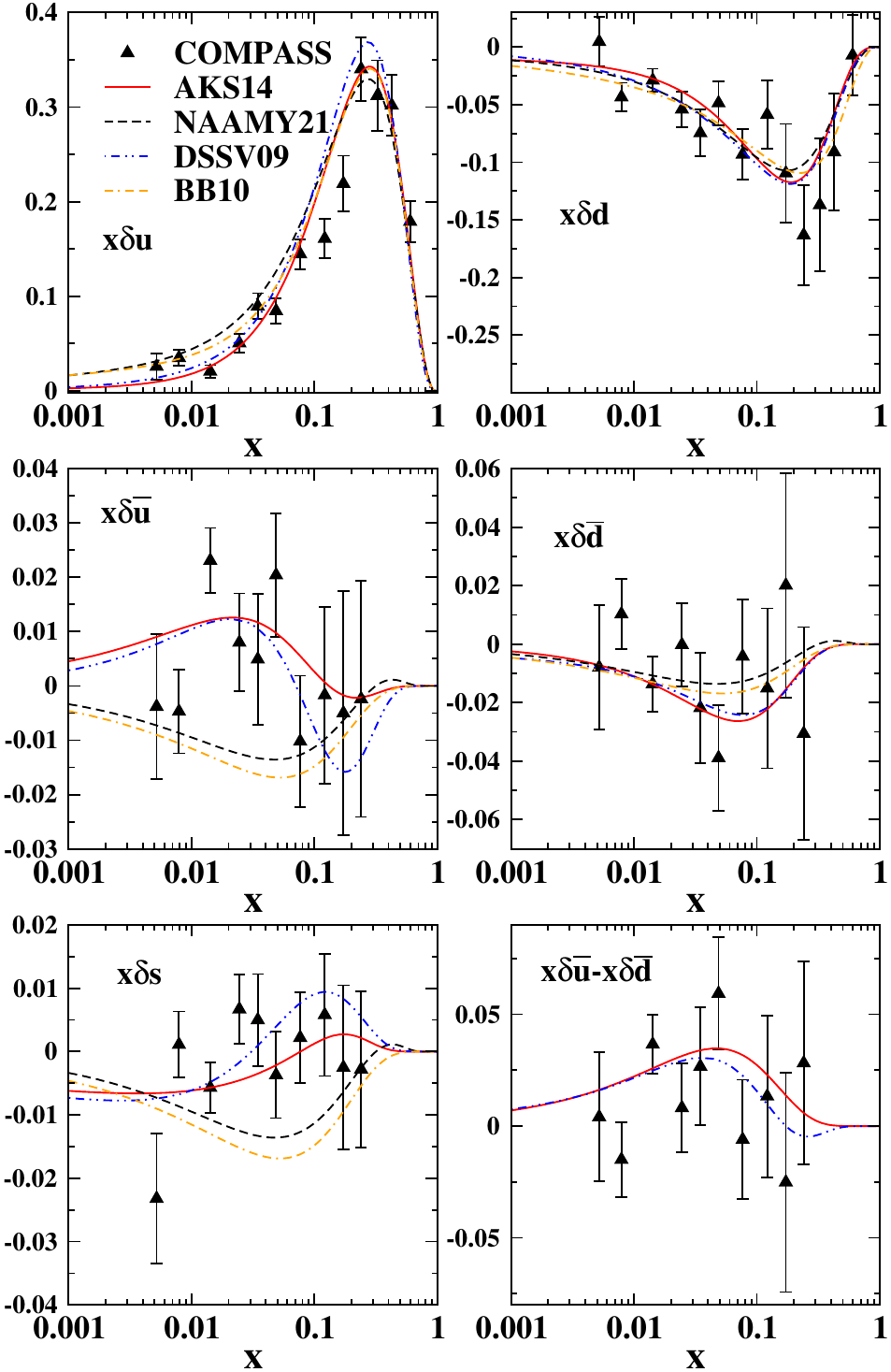}
\begin{center}
\caption{{\small The NLO polarized PDFs from {\tt AKS14}, {\tt NAAMY21},
{\tt DSSV09}~\cite{deFlorian:2009vb} and {\tt BB10}~\cite{Blumlein:2010rn}  in
comparison with COMPASS~\cite{COMPASS:2010hwr} experimental data at $Q^2=3  \,  \textrm{GeV}^2$. \label{fig:figpdfcomp}}}
\end{center}
\end{figure}

\begin{figure}[!htb]
\includegraphics[clip,width=0.45\textwidth]{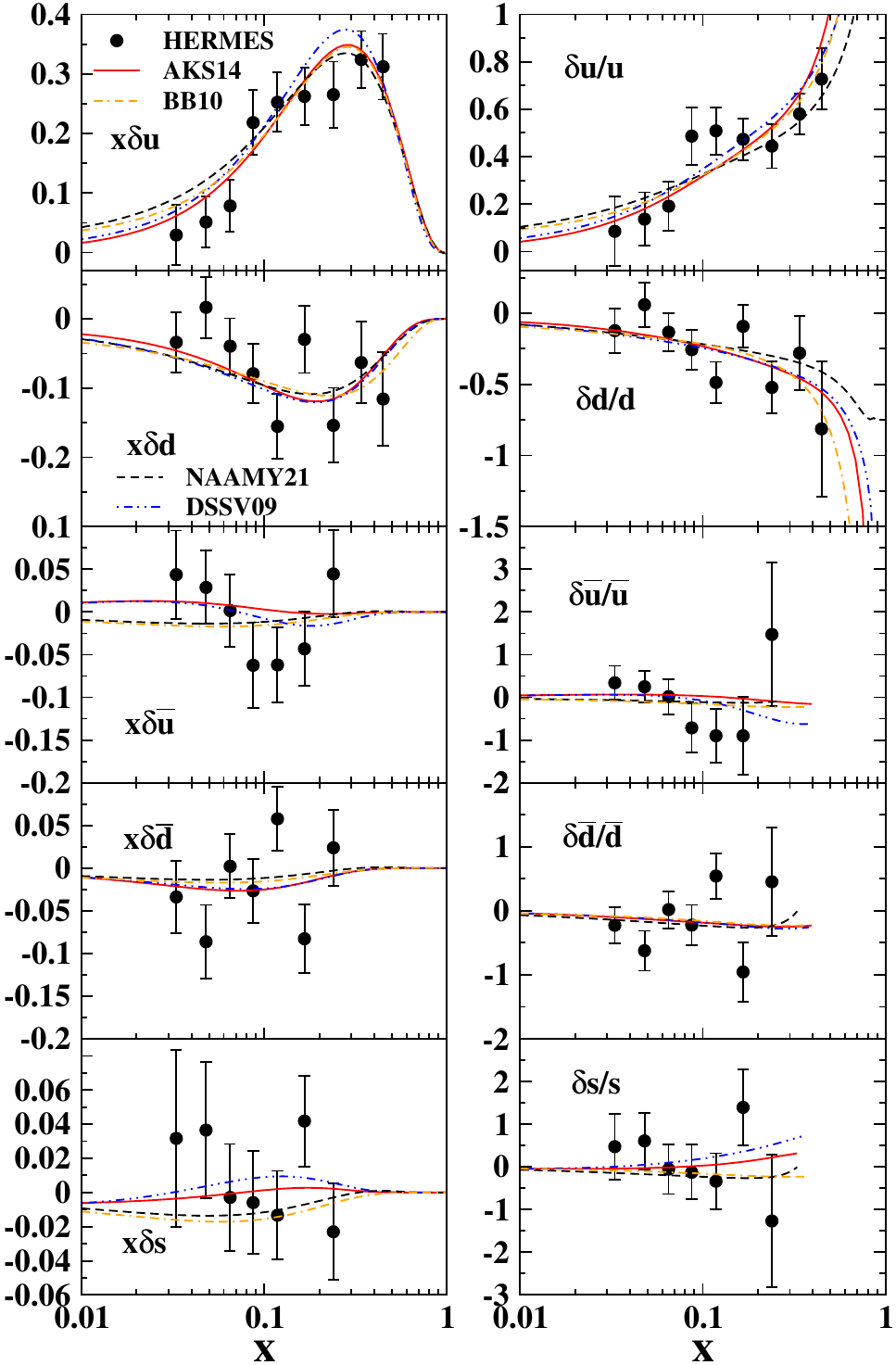}
\begin{center}
\caption{{\small Same as Fig.~\ref{fig:figpdfcomp} but this time in comparison with HERMES experimental
data~\cite{HERMES:2004zsh} at $Q^2=2.5 \, \textrm {GeV}^2$.  \label{fig:figpdfher}}}
\end{center}
\end{figure}

It is worth highlighting once again that the {\tt AKS14} model also 
conducted an additional analysis that included a positive polarized gluon 
to explore the impact of SU(2) and SU(3) symmetry breaking, as well 
as the effect of SIDIS data on the polarized gluon distribution.
In Figs.~\ref{fig:figpdfcomp} and ~\ref{fig:figpdfher} we
show the polarized PDFs obtained from {\tt AKS14} and {\tt NAAMY21} in comparison
with experimental data
from COMPASS~\cite{COMPASS:2010hwr} and HERMES~\cite{HERMES:2004zsh}  experiments.
The results of two other available models, {\tt DSSV09}~\cite{deFlorian:2009vb} and
{\tt BB10}~\cite{Blumlein:2010rn}, are also shown for comparison.
To calculate the $\delta q/q$ in Fig~\ref{fig:figpdfher}, the most recent
unpolarized PDFs from {\tt CT18}~\cite{Hou:2019efy} is used.

Considering the results presented in Figures \ref{fig:figpdfcomp} 
and \ref{fig:figpdfher}, several remarks can be made.
As shown in these plots, in general,
the compatibility of the {\tt AKS14} model with the experimental data
and their error is better than those of {\tt NAAMY21},
more especially for the case of $\delta s$.
 However, it is important to note that the compatibility 
 of the data with the {\tt NAAMY21} model is also good.

As can be seen, the {\tt DSSV09} results which come from the analysis with
symmetry breaking consideration are
mostly close to the {\tt AKS14} curves and both are in
agreement with the experimental data.
However, due to the wide dispersion of the related data,
the plots might  not be able to support them as one may expect.
Also it is obvious that the results of {\tt BB10},
with no symmetry breaking, is closer to the {\tt NAAMY21}.
For the case of $x \delta \bar{u} - x \delta \bar{d}$, both the 
{\tt AKS14} and {\tt DSSV09} models show the same level of 
compatibility and are in agreement with the COMPASS data.
For the specific parton species, it is observed that both the 
{\tt AKS14} and {\tt DSSV09} models demonstrate a better agreement 
with the COMPASS data in terms of $x \delta \bar{u}$ 
and $x \delta \bar{d}$ compared to other groups.
Significant deviations from the COMPASS data
are observed for the {\tt NAAMY21} and  {\tt BB10} for the case of
$x \delta \bar{u}$ and $x \delta s$.

\begin{figure}[!htb]
\includegraphics[clip,width=0.4\textwidth]{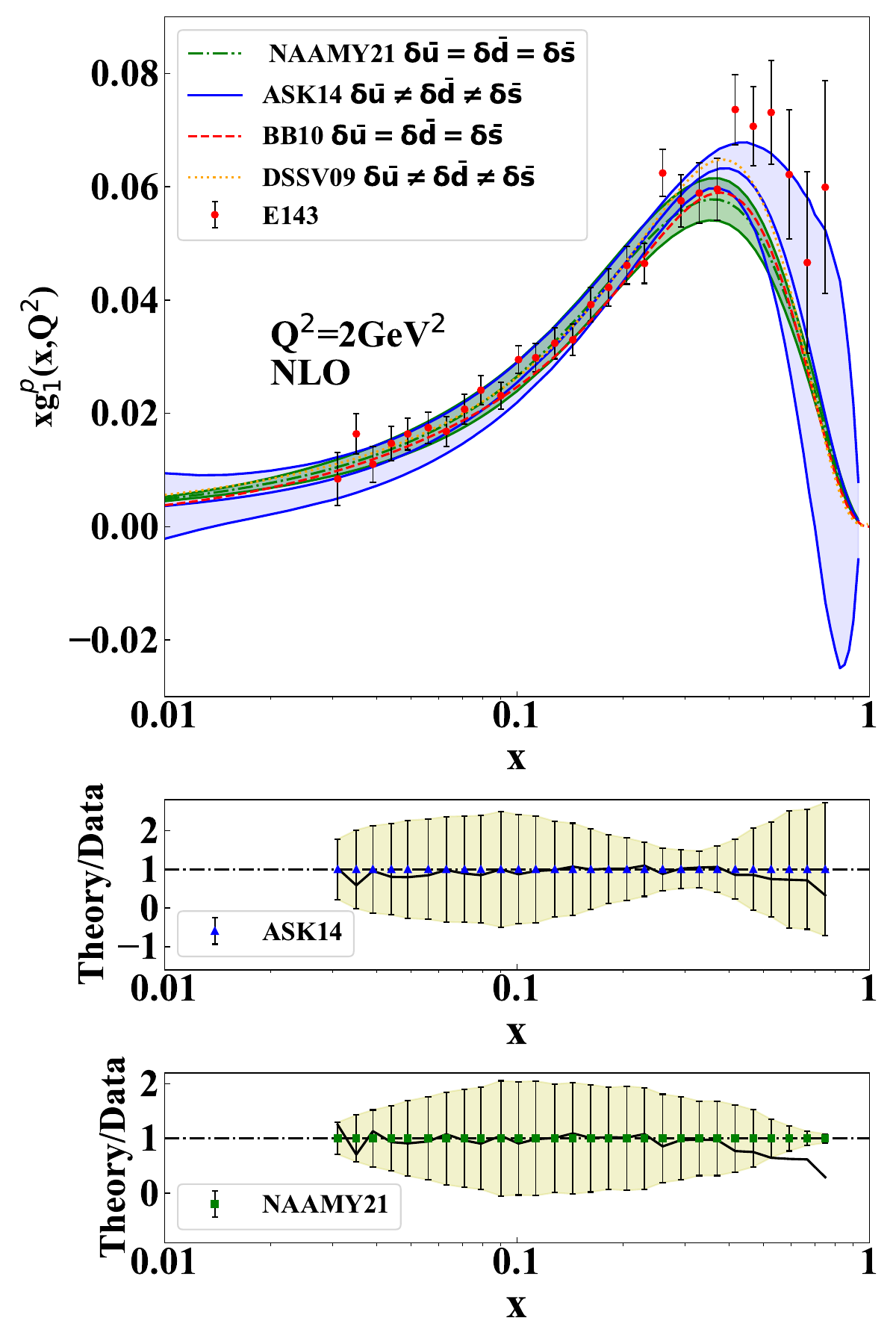}
\begin{center}
\caption{{\small The NLO prediction for the proton polarized
structure function calculated using the
{\tt AKS14}, {\tt NAAMY21},
{\tt DSSV09}, {\tt BB10} in comparison with
E143 experimental data~\cite{E143:1998hbs} at $Q^2=2  \, \textrm {GeV}^2$.  \label{fig:fig3}}}
\end{center}
\end{figure}

\begin{figure}[!htb]
\includegraphics[clip,width=0.4\textwidth]{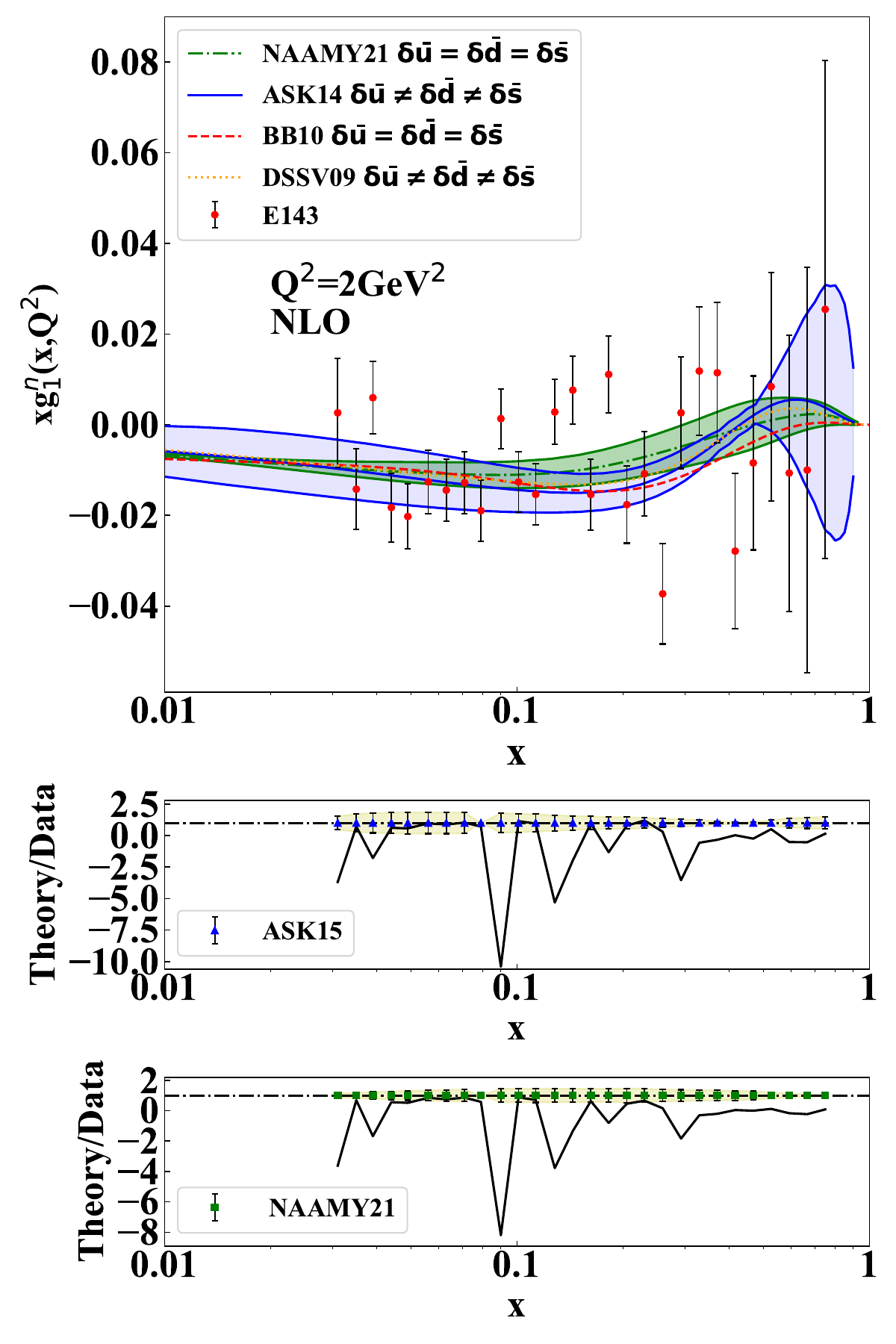}
\begin{center}
\caption{{\small The neutron polarized structure function at $Q^2=2  \, \textrm{GeV}^2$,
compared with {\tt AKS14},
{\tt NAAMY21}, {\tt DSSV09},
{\tt BB10} and E143 experimental data~\cite{E143:1998hbs}.  \label{fig:fig4}}}
\end{center}
\end{figure}

\begin{figure}[!htb]
\includegraphics[clip,width=0.4\textwidth]{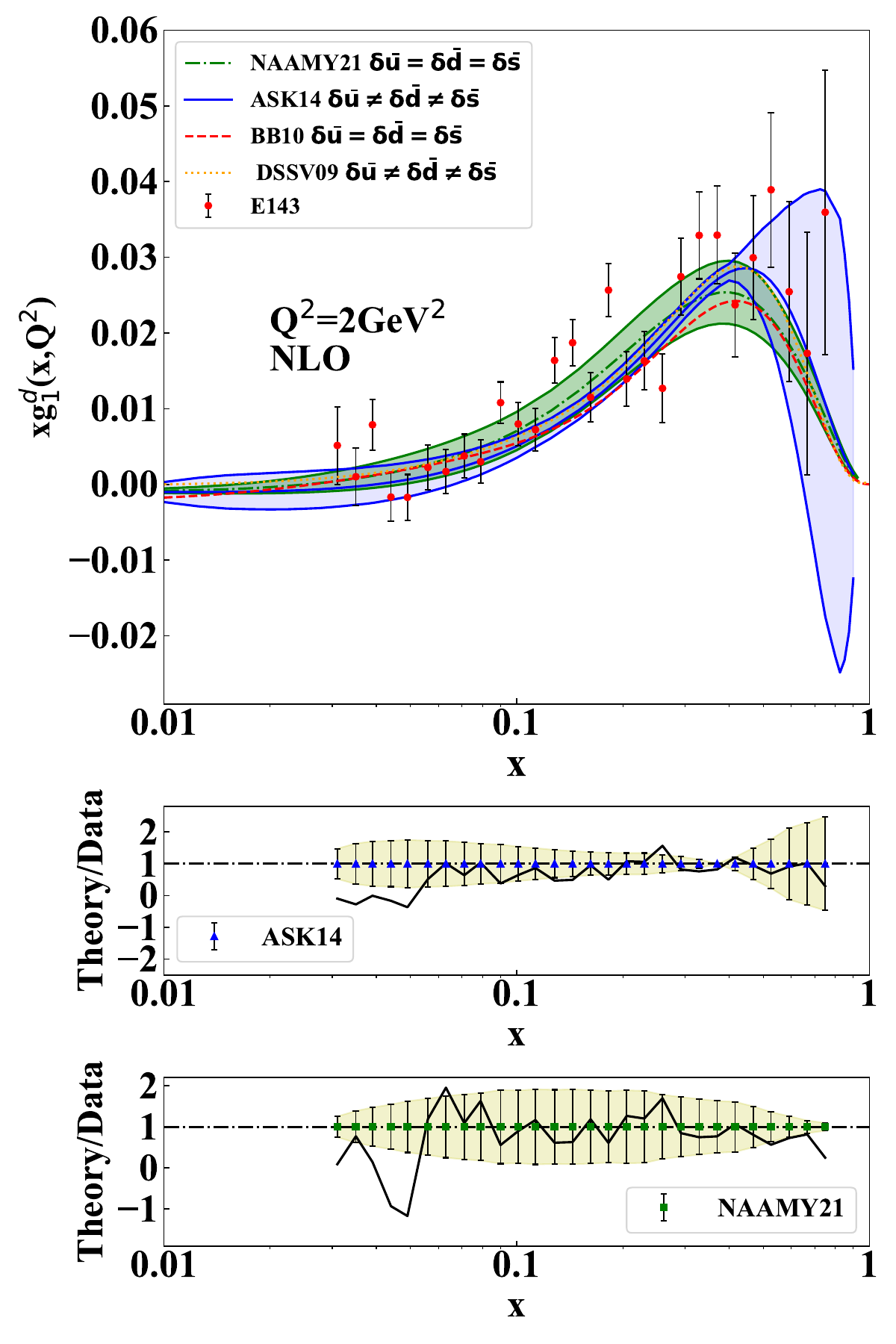}
\begin{center}
\caption{{\small The deuteron polarized structure function at $Q^2=2  \, \textrm{GeV}^2$,
compared with {\tt AKS14},
{\tt NAAMY21}, {\tt DSSV09},
{\tt BB10} and E143 experimental data~\cite{E143:1998hbs}.  \label{fig:fig5}}}
\end{center}
\end{figure}

With the moment expressions for the polarized structure functions of the 
proton, neutron, and deuteron presented in Eqs.\ref{eq1}, \ref{eq2}, and \ref{eq3}, 
respectively, and utilizing the Jacobi polynomials, the structure functions can be calculated in the Bjorken $x$ space.
Figs.~\ref{fig:fig3}, \ref{fig:fig4}, and \ref{fig:fig5} show the comparison of
the structure functions of proton, neutron, and deuteron with
{\tt E143} experimental data~\cite{E143:1998hbs}.
We show both the absolute distributions (upper panel)
and the data/theory ratios (down panel) for
detailed comparison.
In order to correctly assess weather or not the different theoretical results
are statistically different one from the other, and to examine the compatibility
with the data, the theory predictions for different polarized PDFs sets are also displayed
with error bands.
The results from {\tt DSSV09} and {\tt BB10} are also shown and compared.

In general, as observed from these plots, the nucleon polarized structure 
functions generated by the {\tt AKS14} model, which takes into account 
symmetry breaking, are quite close to the {\tt DSSV09} polarized structure functions, 
where SU(2) and SU(3) symmetry breaking is also considered.

Focusing on the comparison with the proton polarized structure function 
shown in Figure \ref{fig:fig3}, one can observe a larger error band for the {\tt AKS14} model. The theoretical prediction of 
the {\tt AKS14} model and the data are compatible over the range of medium to small values of $x$.
Although, differences are also
seen at large $x$, they are always compatible within uncertainties.
As one can see from the data/theory ratios (down panel),
the compatibility of {\tt AKS14} theory prediction and the {\tt E143} experimental data
are  better than those of {\tt NAAMY21} at large $x$. 
The compatibility of the {\tt AKS14} and {\tt NAAMY21} theory predictions 
with the data is similar for small values of $x$.

In Fig.~\ref{fig:fig4}, we present a comparison of the neutron 
polarized structure function with data from the E143 Collaboration. 
Our analysis indicates that both the {\tt AKS14} and {\tt NAAMY21} 
curves demonstrate a good level of agreement with the central points 
of the experimental data. Specifically, for small to medium values of $x$, 
the theoretical predictions of both the {\tt AKS14} and {\tt NAAMY21} 
models exhibit better accordance with the data, considering the uncertainties, 
when compared to other models. However, it is important to note that deviations 
from the experimental data can be observed for 
both models, particularly at large values of $x$.

Finally, focusing on the deuteron polarized structure function in Fig.~\ref{fig:fig5}, 
it can be observed that the NLO prediction of the structure function 
calculated from the {\tt AKS14} model is in good agreement with 
the {\tt E143} experimental data across the entire range of $x$. 
The {\tt DSSV09} polarized structure functions are also seen to 
be close to the {\tt AKS14} curves, as both models account for 
SU(2) and SU(3) symmetry breaking. Comparing with the {\tt NAAMY21} model, 
the data/theory ratios (bottom panel) indicate that the 
compatibility of the {\tt AKS14} theory prediction is better. 
This difference is particularly evident at very large and small 
values of $x$, taking into account the uncertainties.

In Table.~\ref{table3}, we present the value of the
moments of   nucleon polarized structure functions
at $3~{\textrm GeV}^2$ for {\tt AKS14},
{\tt NAAMY21}  polarized PDFs.
Comparing the values of $\Gamma_1^p=0.139\pm0.003\pm0.009\pm0.005$,
$\Gamma_1^n=-0.041\pm0.006\pm0.011\pm0.005$,
and $\Gamma_1^d=0.049\pm0.003\pm0.004\pm0.004$,
reported by COMPASS experimental group~\cite{COMPASS:2015mhb} with
the values reported in Table.~\ref{table3}, it is observed that the values of
moments of neutron and deuteron polarized structure functions, when
symmetry breaking is taken
into account in the {\tt AKS14}, are more compatible with COMPASS results.
 For the case of proton, the value extracted from
{\tt NAAMY21}  is in   better agreement with the COMPASS results.

\begin{table}[ht]
\begin{tabular}{cccc}
\hline\hline
& $\Gamma _{1}^{p}$ & $\Gamma _{1}^{n}$ & $\Gamma _{1}^{d}$ \\
{\tt AKS14} & $0.1273\pm 0.0003$ & $-0.0374\pm 0.00035$ & $0.0406\pm 0.00032$ \\
{\tt NAAMY21} & $0.1322\pm 0.0053$ & $-0.0554\pm 0.0056$ & $0.0350\pm 0.0050$ \\
\hline
\end{tabular}
\caption{The first momentum of the nucleon polarized structure functions calculated using {\tt AKS14} and
{\tt NAAMY21} polarized PDFs.}
\label{table3}
\end{table}

\section{Polarized structure functions of nuclei}\label{psfn}

This section emphasizes the calculation of the polarized structure 
functions of light nuclei, as well as the impact of the symmetry breaking 
of polarized light sea quarks on the agreement between theory and experimental data.
The calculations in the previous section showed that the
polarized PDFs, polarized structure functions of
proton, neutron, deuteron, and their first moments obtained from the
{\tt AKS14} polarized PDFs, in general, have better
compatibility with the experimental data
from COMPASS~\cite{COMPASS:2010hwr,COMPASS:2015mhb}, and E143~\cite{E143:1998hbs}.

In the following section, we analyze the polarized structure functions of light nuclei
Helium-3 $^3$He and Tritium $^3$H.
These two are trivalent light nuclei that, respectively, consist of two protons
plus one neutron and two neutrons plus one proton.
Despite their wave function at ground state level is in the $S$ state,
the higher states $S^\prime$ and $D$ can be also found in a more realistic definition.
At $S^\prime$ and $D$ states, the spin contributions of two protons in Helium-3 or the
two neutrons in Tritium are not canceled and must be considered in
the calculations. The calculation of the polarized structure function can be
performed using the contribution of proton and neutron
polarized structure function in
addition to the spin-dependent nucleon light-cone momentum
distributions $\Delta f^p_{^3{\mathrm He}}$ 
and $\Delta f^n_{^3{\mathrm He}}$~\cite{Yazdanpanah:2009zz,CiofidegliAtti:1993zs,Bissey:2000ed,Bissey:2001cw,Afnan:2003vh}.
They are given by,

\begin{eqnarray}\label{g13H}
	g_1^{^3{\mathrm He}} = &&  \int_x^3 \frac{dy}{y} \Delta f^n_{^3{\mathrm He}}(y) g_1^n(x/y) \\ \nonumber
	&&+2 \int_x^3 \frac{dy}{y} \Delta f^p_{^3{\mathrm He}}(y) g_1^p(x/y) \,,  \nonumber
\end{eqnarray}

\begin{eqnarray}\label{g13H}
	g_1^{^3{\rm H}} = &&  \int_x^3 \frac{dy}{y} \Delta f^n_{^3{\mathrm He}}(y)  g_1^p (x/y) \\ \nonumber
	&&+2 \int_x^3 \frac{dy}{y} \Delta f^p_{^3{\mathrm He}}(y) g_1^n (x/y)   \,.\nonumber
\end{eqnarray}

\begin{figure}[!htb]
\includegraphics[clip,width=0.4\textwidth]{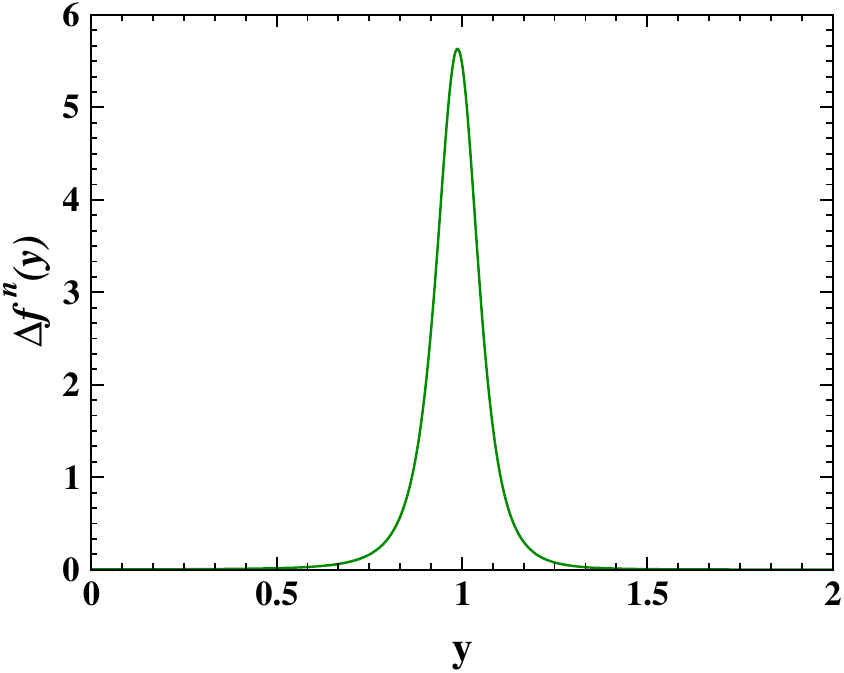} \\
\includegraphics[clip,width=0.4\textwidth]{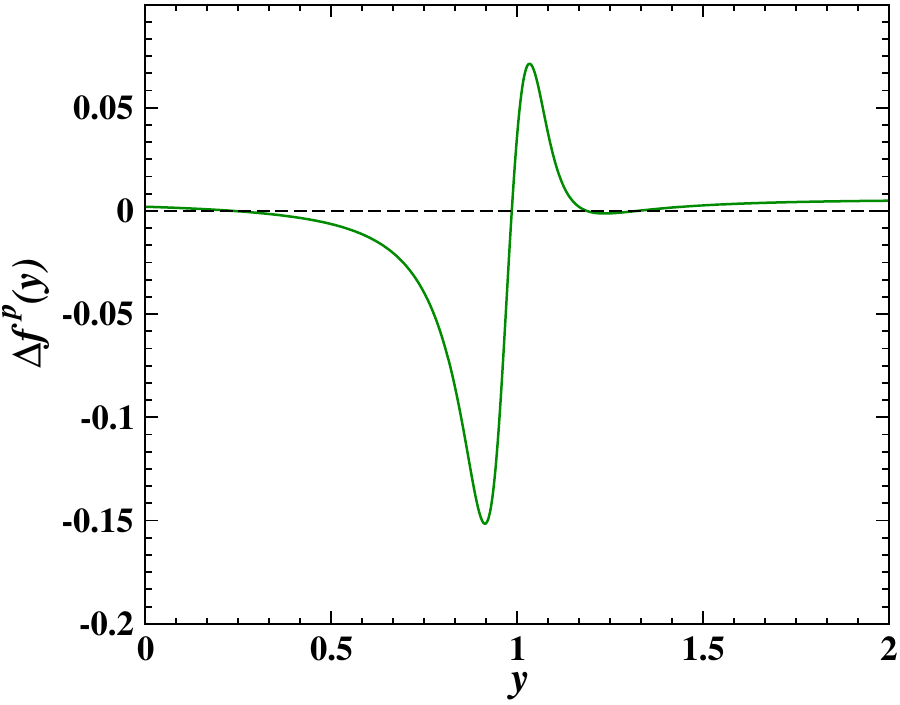}
\begin{center}
\caption{{\small The Spin-dependent nucleon light-cone momentum
distributions $\Delta f^p_{^3{\mathrm He}}$ and
$\Delta f^n_{^3{\mathrm He}}$~\cite{Yazdanpanah:2009zz} as a function of $y$.
\label{fig:fig8}}}
\end{center}
\end{figure}

It is noticeable that because of isospin symmetry, the light cone
momentum distribution $\Delta f^p_{^3{\mathrm He}}$ and
$\Delta f^n_{^3{\mathrm H}}$ are equal in size.
In Fig.~\ref{fig:fig8} we show the spin-dependent nucleon light-cone momentum
distributions  $\Delta f^n_{^3{\mathrm He}}$
and $\Delta f^p_{^3{\mathrm He}}$~\cite{Yazdanpanah:2009zz}  as a 
function of $y$. 
For the calculation of  
$\Delta f^n_{^3{\mathrm He}}$
and $\Delta f^p_{^3{\mathrm He}}$, we utilized the PEST nucleon-nucleon (N-N) interaction 
which is derived from the Paris N-N potential~\cite{Lacombe:1980dr}.

In addition to the correction discussed above, various calculations show that
the main contribution
to the $g_1$ should originate from the $n \rightarrow \Delta^{\circ}$
non-diagonal transition in $^3$He and
from the $p\rightarrow\Delta^{+}$ non-diagonal transition
in $^3$H case~\cite{Frankfurt:1996nf,Saito:1990aj,Carlson:1991ju,Boros:2000af}.

\begin{eqnarray}
	\label{g13Hetri}
	&&g_1^{^3{\rm He}}(x,Q^2)  = \int_x^3 \frac{dy}{y}
	\Delta f^n_{^3{\mathrm He}}(y)g_1^n(x/y, Q^2)  +   \nonumber   \\
	&&2 \int_x^3 \frac{dy}{y} \Delta f^p_{^3{\mathrm He}}(y)g_1^p(x/y, Q^2)\nonumber\\ &+&	4P_{p\rightarrow\Delta^{+}}g_1^{p\rightarrow\Delta^{+}}(x,Q^2)+2P_{n\rightarrow
		\Delta^{\circ}}g_1^{n\rightarrow
		\Delta^{\circ}}(x,Q^2) \,,   \nonumber   \\
\end{eqnarray}

\begin{eqnarray}
	\label{g13Htri}
	&&g_1^{^3{\rm H}}(x,Q^2) =  \int_x^3 \frac{dy}{y}
	\Delta f^n_{^3{\mathrm He}}(y)g_1^p(x/y, Q^2)  +   \nonumber   \\
	&&2\int_x^3 \frac{dy}{y} \Delta f^p_{^3{\mathrm He}}(y)g_1^n(x/y,Q^2)\nonumber\\ &-&	2P_{n\rightarrow
		\Delta^{\circ}}g_1^{p\rightarrow\Delta^{+}}(x,Q^2)-4P_{p\rightarrow\Delta^{+}}g_1^{n\rightarrow
		\Delta^{\circ}}(x,Q^2) \,,  \nonumber   \\
\end{eqnarray}

As a result, the structure functions $g_1^{n\rightarrow\Delta^{\circ}}$
and $g_1^{p\rightarrow\Delta^{+}}$ should be considered in
the calculations, which will lead to a correction for the three
particle nuclei structure function.
Regarding to the model-independent equation~\cite{Bissey:2001cw}, we have

\begin{equation}\label{eq16}
	g_1^{p\rightarrow\Delta^{+}}(x,Q^2)=g_1^{n\rightarrow
		\Delta^{\circ}}(x,Q^2)=\frac{2\sqrt{2}}{5}(g_1^p-4g_1^n),
\end{equation}

and the polarized structure function equation will be written as,

\begin{eqnarray}\label{g_1He-degrees-of-freedom}
	&&g_1^{^3{\mathrm He}}=\int_x^3 \frac{dy}{y} \Delta f^n_{^3{\mathrm He}}(y) g_1^n(x/y)
	+2 \int_x^3 \frac{dy}{y} \Delta f^p_{^3{\mathrm He}}(y) g_1^p(x/y)      \nonumber  \\
	&& -0.014 \Big(g_1^p(x) - 4 g_1^n(x) \Big) \,,
\end{eqnarray}

and

\begin{eqnarray}\label{g_1H-degrees-of-freedom}
	&&g_1^{^3{\rm H}} =  \int_x^3 \frac{dy}{y} \Delta f^n_{^3{\mathrm He}}(y)  g_1^p (x/y)
	+2 \int_x^3 \frac{dy}{y} \Delta f^p_{^3{\mathrm He}}(y) g_1^n (x/y)   \nonumber \\
	&& + 0.014 \Big(g_1^p (x) - 4 g_1^n (x)  \Big)  \,.
\end{eqnarray}

At high energy levels or at low values of Bjorken $x$,
the virtual photon can interact with multiple nucleons coherently.
This behavior is apparent in nuclear targets.
Nuclear shadowing
and antishadowing are examples of these coherent
effects~\cite{Bissey:2001cw,Ethier:2013hna}. According to the shadowing
and antishadowing correction,
the Helium-3 and Tritium nuclei polarized structure functions can be rewritten as,

\begin{eqnarray}\label{g1Hesh}
	&& g_1^{^3{\mathrm He}}=\int_x^3 \frac{dy}{y}
	\Delta f^n_{^3{\mathrm He}}(y) g_1^n(x/y)
	+2 \int_x^3 \frac{dy}{y} \Delta f^p_{^3{\mathrm He}}
	(y) g_1^p(x/y)  \nonumber \\
	&& - 0.014 \Big( g_1^p(x) - 4 g_1^n(x)\Big) +
	a(x) g_1^n(x) + b(x) g_1^p(x)  \ ,\nonumber\\
\end{eqnarray}

and

\begin{eqnarray}\label{g1Hsh}
	&&g_1^{^3{\mathrm H}} =  \int_x^3 \frac{dy}{y}
	\Delta f^n_{^3{\mathrm He}}(y) g_1^p(x/y)
	+ 2\int_x^3 \frac{dy}{y} \Delta f^p_{^3{\mathrm He}}(y)
	g_1^n (x/y)  \nonumber \\
	&& + 0.014 \Big( g_1^p(x) - 4 g_1^n (x) \Big) +
	a(x) g_1^n(x) + b(x) g_1^p(x)  \,, \nonumber\\
\end{eqnarray}

where $a(x)$ and $b(x)$ functions describe the shadowing and
anti-shadowing effect as a function of $x$ and $Q^2$~\cite{Bissey:2001cw}.
Since the current experimental data do not cover very small
values of $x$ considerably, the corrections from
shadowing $(10^{-4}\leq x\leq 0.03-0.07)$ and $(0.03-0.07\leq x \leq 0.2)$ anti-shadowing
can be completely ignored in the calculations of 
polarized nuclei~\cite{JeffersonLabHallA:2016neg}.
However, the calculations of Refs.~\cite{Bissey:2001cw,Ethier:2013hna}
have shown that these effects are quite significant and
impact the extraction of the nucleon polarized
structure functions at small values of Bjorken $x$.

\begin{figure}[!htb]
\includegraphics[clip,width=0.45\textwidth]{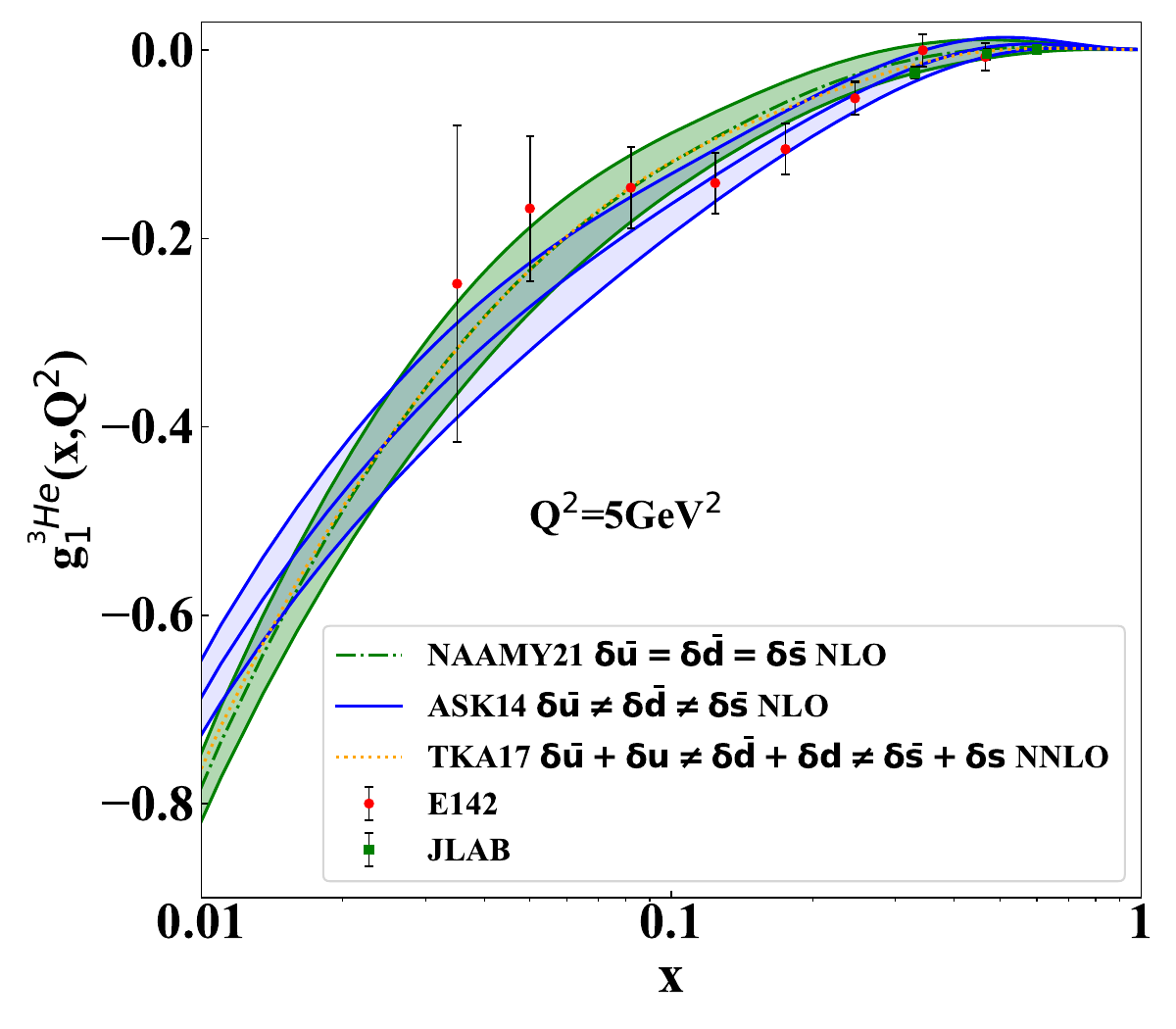}
\begin{center}
\caption{{\small The theory prediction for $g_1^{^3\rm{He}}$ polarized structure functions
at NLO and NNLO approximation from {\tt NAAMY21}, {\tt AKS14} and {\tt KTA17} in comparison
with experimental data from {\tt E142}~\cite{E142:1996thl}
and {\tt JLAB}~\cite{JeffersonLabHallA:2004tea}. \label{fig:fig9}}}
\end{center}
\end{figure}

\begin{figure}[!htb]
\includegraphics[clip,width=0.45\textwidth]{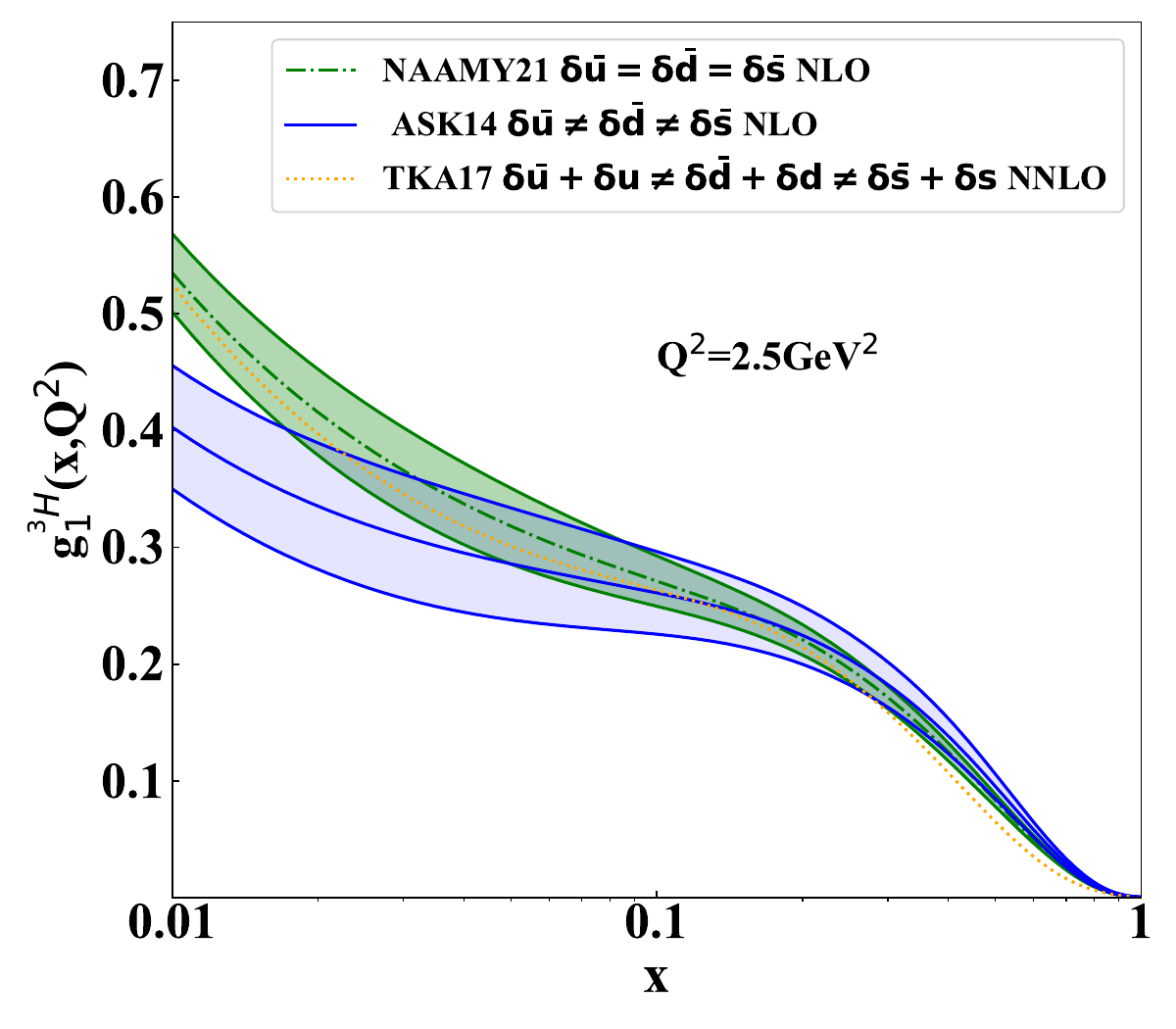}
\begin{center}
\caption{{\small Same as Fig.~\ref{fig:fig9} but this time for
the $g_1^{^3\rm{H}}$ polarized structure functions.\label{fig:fig10}}}
\end{center}
\end{figure}

According to Eqs.~\ref{g_1He-degrees-of-freedom} and~\ref{g_1H-degrees-of-freedom},
the Helium-3 and Tritium polarized structure functions are
extracted at $5 \, \textrm{GeV}^2$ and $2.5 \, \textrm{GeV}^2$ respectively and
they are shown in Figs.~\ref{fig:fig9} and~\ref{fig:fig10}.

Polarized structure functions of $^3$He and $^3$H at NLO and NNLO from
{\tt NAAMY21} and
{\tt KTA17}~\cite{Khanpour:2017fey} polarized PDFs, both concerning symmetry
of light sea quarks, are compared with {\tt AKS14}~\cite{Arbabifar:2013tma} model at
NLO concerning symmetry breaking of light sea quarks.

In Fig.\ref{fig:fig9}, we also 
include the {\tt E142}\cite{E142:1996thl} and 
JLAB~\cite{JeffersonLabHallA:2004tea} experimental data 
for comparison. As shown in this figure, the theory 
prediction from the {\tt AKS14} model, even though it is 
similar to that of {\tt NAAMY21}, exhibits slightly better 
compatibility with the experimental results across most of 
the Bjorken $x$ region covered by the data. This compatibility 
is particularly emphasized for larger 
values of $x$ ($x > 0.1$) within the uncertainty bands.

In Fig.~\ref{fig:fig10} and for the case of $g_1^{^3\rm{H}}$ polarized structure functions,
one can see that the difference between the curves with
and without considering symmetry breaking is noticeable,
especially at low value of $x$.

\begin{figure}[!htb]
\includegraphics[clip,width=0.45\textwidth]{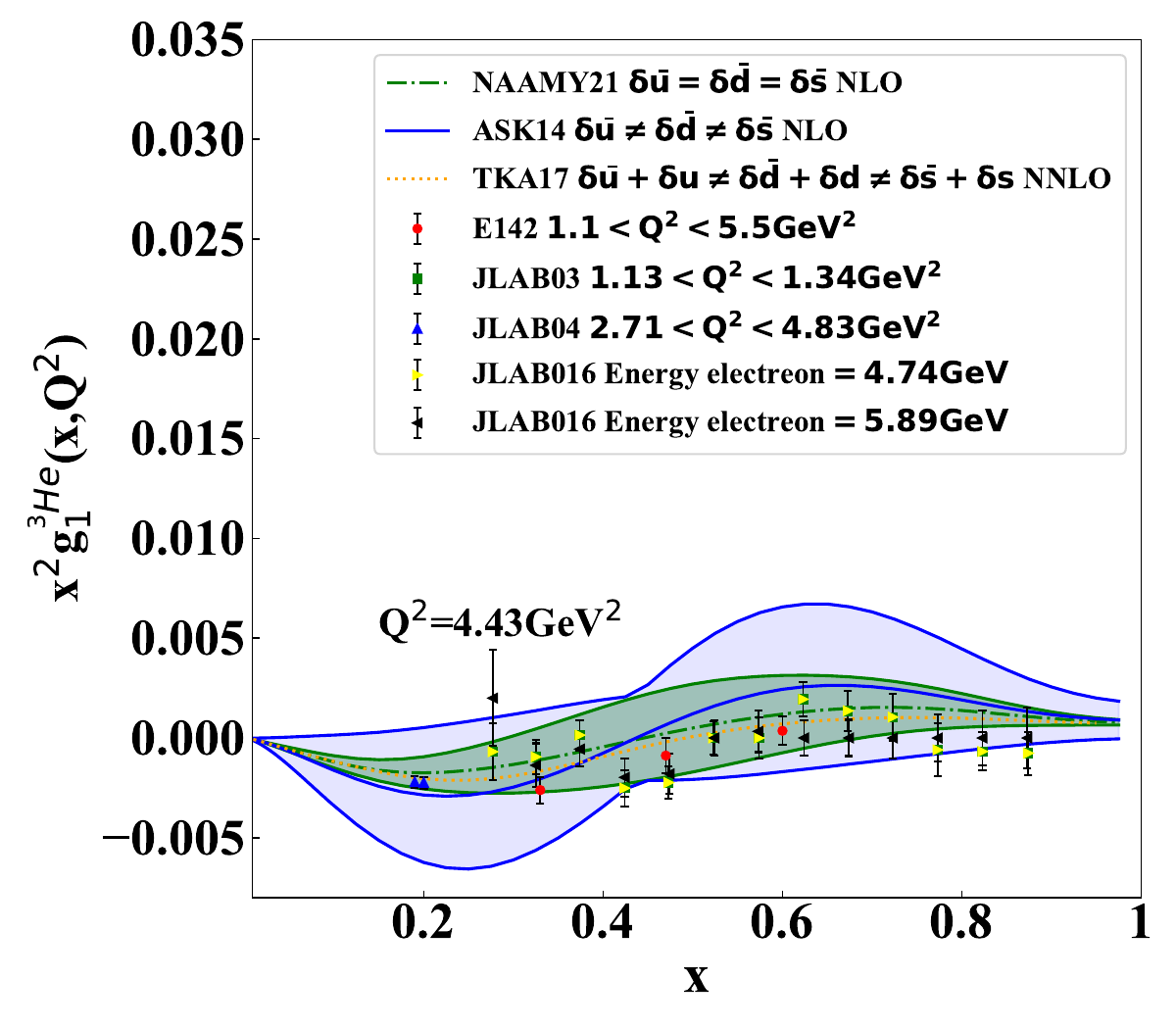}
\begin{center}
\caption{{\small Same as Fig.~\ref{fig:fig9} but this time for
the $x^2g_1^{^3\rm{He}}$ polarized structure functions
in  comparison with experimental data from
{\tt E142}~\cite{E142:1996thl}, {\tt JLAB04}~\cite{JeffersonLabHallA:2004tea},
{\tt JLAB03}~\cite{jlab:WilliamandMary} and
{\tt JLAB16}~\cite{JeffersonLabHallA:2016neg}. \label{fig:fig11}}}
\end{center}
\end{figure}

Fig.~\ref{fig:fig11} shows the comparison of $x^2g_1^{^3\rm{He}}$ polarized
structure function in comparison with
{\tt E142}~\cite{E142:1996thl}, {\tt JLAB04}~\cite{JeffersonLabHallA:2004tea},
{\tt JLAB03}~\cite{jlab:WilliamandMary}, and {\tt JLAB16}~\cite{JeffersonLabHallA:2016neg}
experimental data at different energies in the range of $1.1<\rm{Q}^2<5.89 \, \rm{GeV}^2$.
Based on this plot, it can be concluded that the {\tt  AKS14} model 
exhibits comparable agreement to the other models within the region of $0.1 \leq x \leq 0.4$. 
For higher values of $x$, 
the {\tt NAAMY21} and {\tt KTA17} models show better 
compatibility with the experimental data points. However, 
it should be noted that both the {\tt AKS14} and {\tt NAAMY21} 
models remain compatible within the given 
uncertainties throughout the entire range.

\begin{figure}[!htb]
	\includegraphics[clip,width=0.45\textwidth]{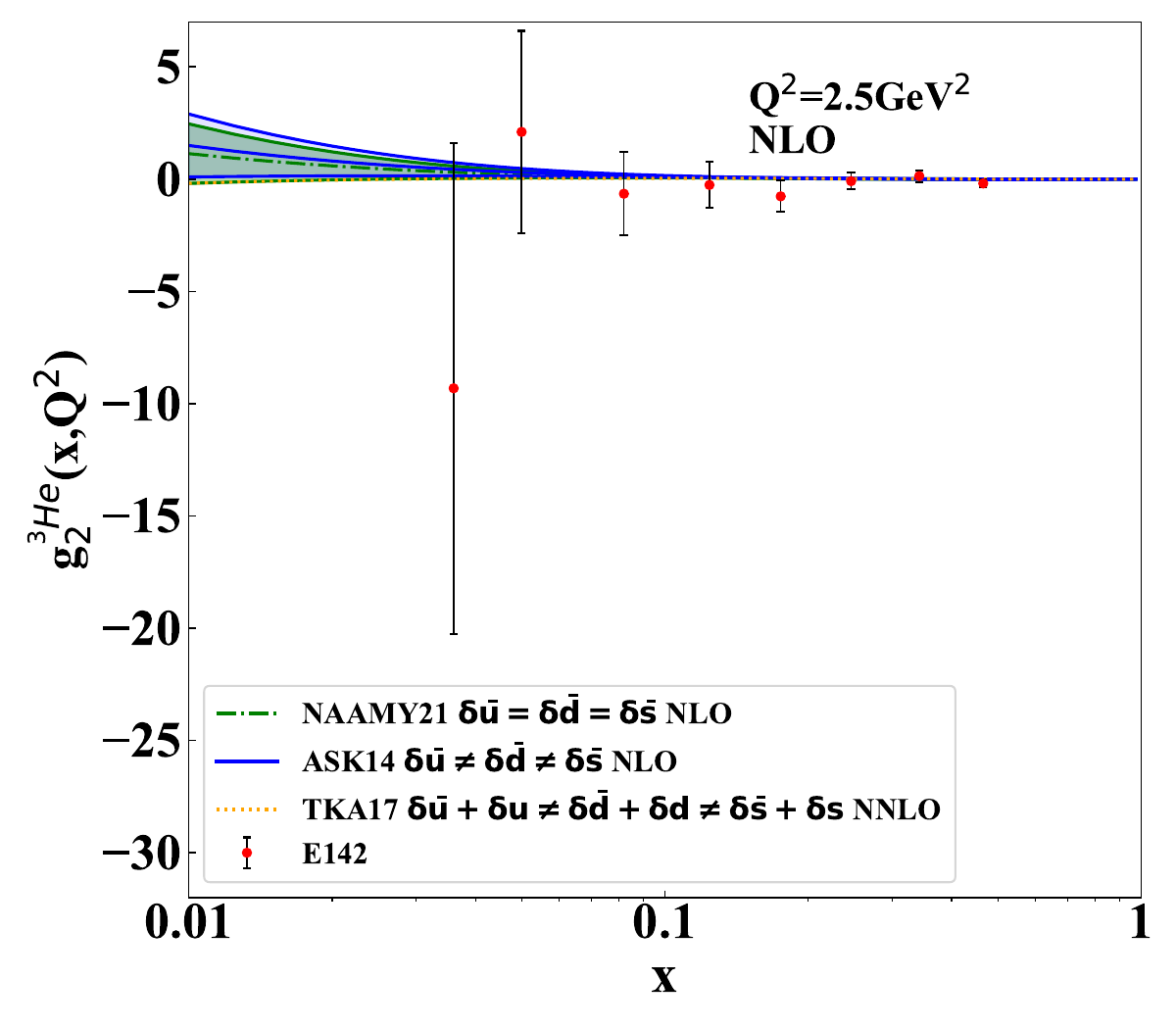}
	\begin{center}
\caption{{\small Same as Fig.~\ref{fig:fig9} but this time for the $g_2^{^3\rm{He}}$
polarized structure functions in comparison
with {\tt E142} experimental data~\cite{E142:1996thl}. \label{fig:fig12}}}
\end{center}
\end{figure}

\begin{figure}[!htb]
\includegraphics[clip,width=0.45\textwidth]{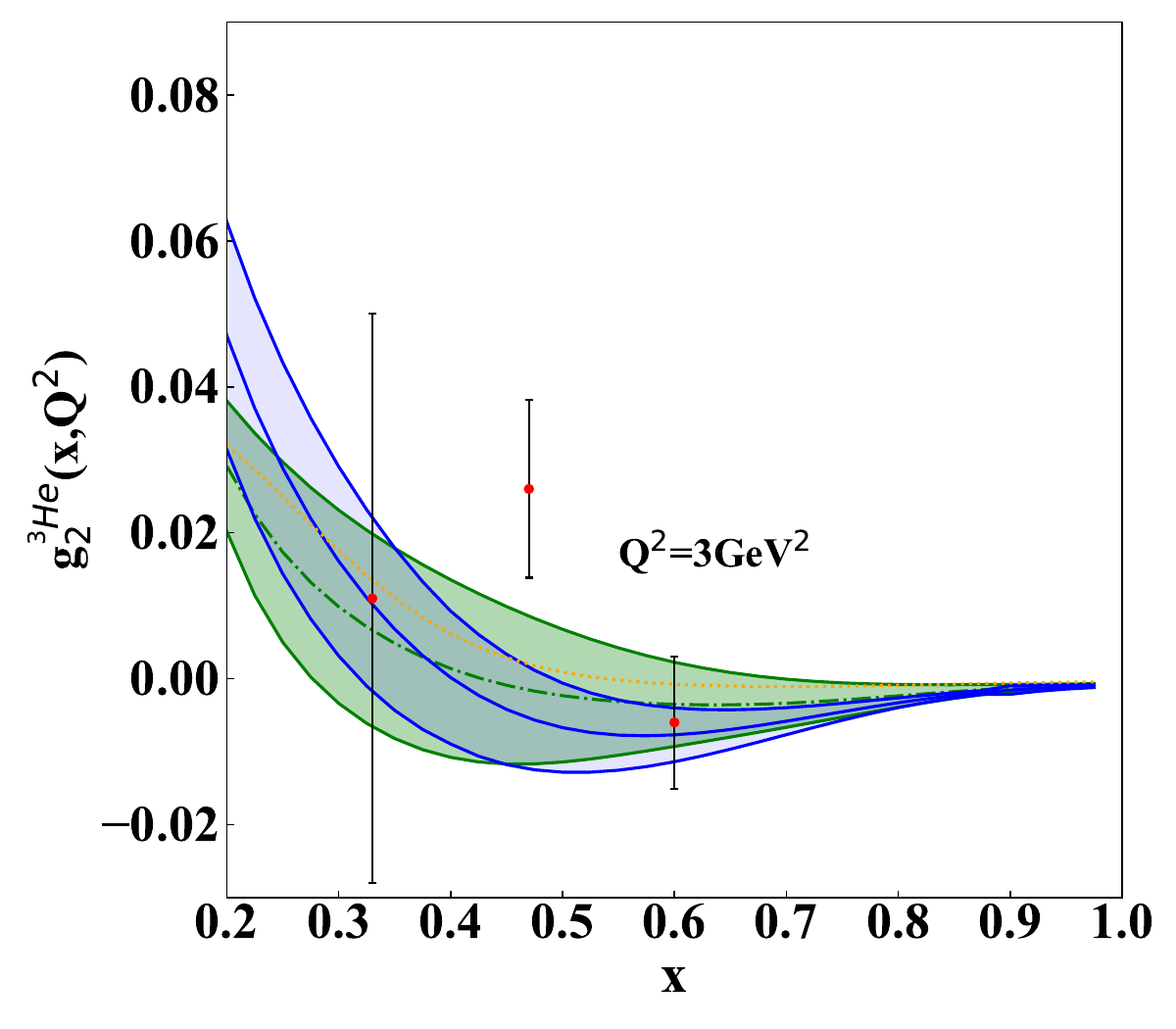}
\begin{center}
\caption{{\small Same as Fig.~\ref{fig:fig9} but this time for the
$g_2^{^3He}$ polarized structure functions in comparison with the
{\tt JLAB} experimental data~\cite{JeffersonLabHallA:2004tea}. \label{fig:fig13}}}
\end{center}
\end{figure}

The Wandzura-Wilczek $g_2$ structure function~\cite{Wandzura:1977qf} can
be calculated as $g_2^{WW}(x)=-g_1(x)-\int_{x}^{1}$ $\frac{dy}{y}g_1(y)$, 
 and for the case of Helium-3 and Tritium structure functions, 
they are given by:

\begin{eqnarray}
&& g_2^{^3{\rm He}} = \int_x^3 \frac{dy}{y} \Delta f^n_{^3{\rm He}}(y) g_2^n(x/y)   \nonumber \\
&&+ 2 \int_x^3 \frac{dy}{y} \Delta f^p_{^3{\rm He}}(y) g_2^p (x/y)  \nonumber \\
&& -0.014 \Big( g_2^p(x) - 4 g_2^n (x) \Big) \,,
\end{eqnarray}

and

\begin{eqnarray}
&&g_2^{^3{\rm H}} =  \int_x^3 \frac{dy}{y} \Delta f^n_{^3{\rm He}}(y) g_2^p (x/y)   \nonumber \\
&&+ 2\int_x^3 \frac{dy}{y} \Delta f^p_{^3{\rm He}}(y) g_2^n (x/y)  \nonumber \\
&& + 0.014 \Big( g_2^p(x) - 4 g_2^n (x)\Big)  \,.
\end{eqnarray}

Figs.~\ref{fig:fig12} and \ref{fig:fig13} display the polarized structure
of $g_2^{^3He}$ extracted from {\tt AKS14} and {\tt NAAMY21} at NLO, and
{\tt KTA17}~\cite{Khanpour:2017fey} at the NNLO approximation
compared to the {\tt JLAB04}~\cite{JeffersonLabHallA:2003joy} and
{\tt JLAB16}~\cite{JeffersonLabHallA:2016neg} experimental data.
Although the curves are not exactly the same, a similar agreement
in the behavior of three models in comparison with
the experimental data of {\tt JLAB04} and
{\tt JLAB16} and their error bar is observed.

\section{Bjorken sum rule}\label{bjorken}

To quantify the nuclear corrections as expressed in Eqs.~\ref{g_1He-degrees-of-freedom} and
\ref{g_1H-degrees-of-freedom} we use the ratio known as $\eta$.
Bjorken sum rule evaluates the difference between
the first moment between Helium and Tritium polarized structure functions.
It is given by,

\begin{equation}
\label{eq:Bjorken-sum-He}
\int_{0}^{3} [g_1^{^3{\mathrm H}}(x,Q^2)-g_1^{^3\mathrm{He}}(x,Q^2)]dx =
\frac{1}{6} g_A|_{triton} [ 1 + {\cal O} (\frac{\alpha_s}{\pi}) ]\,,
\end{equation}

where $g_A|_{triton}=1.211\pm0.002$~\cite{Budick:1991zb}.
According to the Bjorken sum rule~\cite{Bjorken:1966jh} for proton
and neutron  without nuclei constraint we have

\begin{equation}
\label{eq:Bjorken-sum}
\int_{o}^{1} [g_1^p(x, Q^2) - g_1^n(x, Q^2)] dx =
\frac{1}{6} g_A [ 1 + {\cal O} (\frac{\alpha_s}{\pi}) ]\,,
\end{equation}

where $g_A=1.2670\pm0.0035$~\cite{ParticleDataGroup:2016lqr}.
Finally the ratio $\eta$ can be written as,

\begin{eqnarray}\label{eq:sumratio}
	\eta &\equiv& \frac{g_A|_{triton}}{g_A} =  \nonumber \\
	&& \frac{\int_0^3[g_1^{^3{\mathrm H}}(x,Q^2)-g_1^{^3\mathrm{He}}
		(x,Q^2)]dx} {\int_0^1[g_1^p(x,Q^2)-g_1^n(x,Q^2)]dx} = 0.956 \pm 0.004 \,. \nonumber \\
\end{eqnarray}

In the calculations, we obtained a value of $0.936\pm0.002$ for
{\tt NAAMY21} model and $0.954\pm0.003$ for {\tt AKS14}.
This result indicates that the analysis taking into account symmetry 
breaking produces a value of the ratio $\eta$ that is closer to the 
Bjorken sum rule compared to the analysis that neglects symmetry breaking.

\section{Efremov-Leader-Teryaev (ELT) Sum Rule}\label{elt}

The Efremov-Leader-Teryaev (ELT) sum rule can be derived by integrating 
the valence part of the $g_1$ and $g_2$ structure functions over 
the Bjorken $x$ variable in the limit $m_q \rightarrow 0$ \cite{blum:1997}. 
The ELT sum rule is expressed as follows:

\begin{eqnarray}\label{eq:ELT}
\int_0^1x[g_1^{V}(x) + 2g_2^{V}(x)]dx=0~.
\end{eqnarray}

where $g_{1(2)}^{V}$ denotes the valence quark contributions to the $g_{1(2)}$.
When considering the symmetry of light sea quarks and assuming they 
carry an equal fraction of the spin in protons and neutrons, the 
Efremov-Leader-Teryaev (ELT) sum rule can be written as follows:

\begin{eqnarray}\label{eq:ELTsym}
\int_0^1x[g_1^{p}(x)-g_1^n + 2(g_2^{p}(x) - g_2^n(x)]dx=0~.
\end{eqnarray}

The values obtained $0.01017 \pm 0.00004$ and $-0.030763 \pm 0.0004071$
from {\tt NAAMY21} and {\tt AKS14} respectively.
It seems the consideration of the
light sea quarks symmetry breaking, the ELT sum rule is
derived directly from Eq.~\ref{eq:ELT}.
The value of the left hand side of
above equation is obtained $-0.011\pm 0.008$ from {\tt E155}~\cite{E155:2003}
analysis at Q$^2=5~\rm{GeV}^2$.
The symmetry breaking makes the results negative, however  are of a similar magnitude. 
Also it is concluded that disregarding the symmetry
of $\delta \bar{u}$, $\delta \bar{d}$ and $\delta \bar{s}$ can be effective on the
magnitude of  ELT sum rule to be far  from zero.

\section{Lorentz color force components}\label{lfc}

The force exerted by the probing quark on the struck 
quark in the DIS process, 
perpendicular to the direction of motion, 
is referred to as the Lorentz color force.
It is divided into electric and magnetic components~\cite{Burkardt:2008ps}

\begin{equation}\label{eq:F_E}
F_E^{y,n} = -\frac{M_n^2}{6}(2d_2^n + f_2^n)\,,
\end{equation}

\begin{equation}\label{eq:F_B}
F_B^{y,n} = -\frac{M_n^2}{6}(4d_2^n - f_2^n)\,.
\end{equation}

Here $d_2^n$ is twist-3 matrix element and can be evaluated via

\begin{equation}\label{eq:d2}
d_2(Q^2) = \int_{o}^{1}x^2 [2g_1(x, Q^2) + 3 g_2(x, Q^2)] dx \,,
\end{equation}

and $f_2^n$ is twist-4 matrix element and can be extracted from the
first moment equation of $g_1(x,Q^2)$~\cite{Kataev:1994gd}

\begin{eqnarray}
\label{eq:f2}
\eta_1 &\equiv & \int_{o}^{1}g_1(x,Q^2) dx \\
&=&\mu_2 + \frac{M^2}{9 Q^2}(a_2 + 4 d_2 +4 f_2) + {\cal O} \left(\frac{1}{Q^4}\right) \,.
\end{eqnarray}

The quantity $\mu_2$ is known as twist-2 contribution

\begin{eqnarray}
\label{eq:mu2}
\mu_2(Q^2)=C_{ns}(Q^2)\left(-\frac{1}{12}g_A+\frac{1}{36}
a_8\right)+C_s(Q^2)\Delta\Sigma,\nonumber\\
\end{eqnarray}	

in which $C_{ns}$ and $C_s$ are the nonsinglet and singlet Wilson
coefficients~\cite{Larin:1997qq}.
$g_A$ and $a_8$  are respectively
the flavor-triplet and the octet axial charge and $\Delta\Sigma$ denotes
the singlet axial current~\cite{Osipenko:2004xg,Meziani:2004ne}.
The quantity $a_2$ is the third moment of $g_1$ known as $a_2=\int x^2 g_1dx$.

In Table.~\ref{twists}, we present the calculated values for the $d_2^n$, $f_2^n$ and $a_2^n$
using two models of {\tt AKS14} and {\tt NAAMY21}.
The corresponding value from JLAB experiment~\cite{JeffersonLabHallA:2016neg}
also has been shown as well.

Table.~\ref{Lorentzforce} presents the calculated magnetic
and electric Lorentz color force components calculated from {\tt AKS14}
and {\tt NAAMY21} models. The measured values from {\tt JLAB}
for Lorentz color force components~\cite{JeffersonLabHallA:2016neg} also presented in
Table.~\ref{Lorentzforce} for comparison.

As one can see, the magnitudes of $F_{B}^{y,n}$ and $F_{E}^{y,n}$ extracted 
from the {\tt AKS14} QCD analysis, which takes into account the symmetry 
breaking of light sea quarks, exhibit good agreement with the 
measured values from the {\tt JLAB} experiment.
In addition, the calculated Lorentz color forces denoted by $F_{B}^{y,n}$
and $F_{E}^{y,n}$ are in same size with a different signs
as expected from QCD prediction~\cite{Burkardt:2008ps}.
The magnitudes of these two forces, extracted from {\tt NAAMY21} QCD analysis
with no symmetry breaking, are one order of magnitude smaller than {\tt JLAB} measured values and the
{\tt AKS14} prediction as well.
We should mention here that the agreement in the order of magnitude between 
the experimental data and the {\tt AKS14} model arises from a combination of  ingredients 
in which the experimental data and the {\tt AKS14}  model differ significantly.

\begin{table*}[ht]
\begin{tabular}{ccccc}
\hline\hline
& $<Q^{2}>\left[ \textrm{GeV}^{2}\right] $ & {\tt JLAB16}~\cite{JeffersonLabHallA:2016neg} & {\tt NAAMY21}~\cite{Nematollahi:2021ynm} & {\tt AKS14}~\cite{Arbabifar:2013tma}  \\ \hline
$d_{2}^{n}\left[ \times 10^{-5}\right] $ & 3.21 & $-421.0\pm 79.0\pm 82.0\pm
8.0$ & $-125.566\pm 0.1712$ & $-89.488\pm 2.333$ \\
$d_{2}^{n}\left[ \times 10^{-5}\right] $ & 4.32 & $-35.0\pm 83.0\pm 69.0\pm
7.0$ & $-124.783\pm 4.809$ & $-86.225\pm 1.806$ \\ \hline \hline
$f_{2}^{n}\left[ \times 10^{-3}\right] $ & 3.21 & $53.41\pm 0.79\pm 25.55$ &
$9.972\pm 2.166$ & $148.330\pm 25.851$ \\
$f_{2}^{n}\left[ \times 10^{-3}\right] $ & 4.32 & $49.66\pm 0.83\pm 25.99$ &
$30.732\pm 28.246$ & $219.473\pm 35.044$ \\ \hline \hline
$a_{2}^{n}\left[ \times 10^{-4}\right] $ & 3.21 & $8.552\pm 1.761\pm 6.125$
& $1.174\pm 0.01$ & $3.953\pm 1.811$ \\
$a_{2}^{n}\left[ \times 10^{-4}\right] $ & 4.32 & $5.044\pm 2.270\pm 6.042$
& $1.632\pm 0.0003$ & $3.754\pm 1.712$ \\ \hline \hline
\end{tabular}
\caption{The values of $d^n_2$, $a_2^n$ and $f_2^n$ from
{\tt JLAB}~\cite{JeffersonLabHallA:2016neg} experimental data with
statistical and systematic uncertainties. The values calculated by
{\tt AKS14} and {\tt NAAMY21}
polarized PDFs are also shown as well.}\label{twists}
\end{table*}

\begin{table*}[ht]
\begin{tabular}{cccc}
\hline\hline
& $<Q^{2}>\left[ \textrm{GeV}^{2}\right] $ & $F_{E}^{y,n}\left[ \textrm{MeV/fm}%
\right] $ & $F_{B}^{y,n}\left[ \textrm{MeV/fm}\right] $ \\
{\tt JLAB16}~\cite{JeffersonLabHallA:2016neg} & $3.21$ & $-33.53\pm 1.32\pm 19.07$ & $52.35\pm 2.43\pm 19.18$ \\
{\tt JLAB16}~\cite{JeffersonLabHallA:2016neg} & $4.32$ & $-36.48\pm 1.38\pm 19.38$ & $38.04\pm 2.55\pm 19.46$ \\
\hline \hline
{\tt NAAMY21} & $3.21$ & $-3.992\pm 3.264$ & $2.207\pm 6.049$ \\
{\tt NAAMY21} & $4.32$ & $-4.154\pm 6.332$ & $5.256\pm 6.290$ \\ \hline \hline
{\tt AKS14} & $3.21$ & $-21.560\pm 3.796$ & $22.350\pm 3.817$ \\
{\tt AKS14} & $4.32$ & $-32.037\pm 5.150$ & $32.798\pm 5.166$ \\ \hline \hline
\end{tabular}
\caption{Measured magnetic and electric Lorentz color force
components from {\tt JLAB}~\cite{JeffersonLabHallA:2016neg} experiment
along with statistical and systematic uncertainty.
The calculated quantities using {\tt AKS14} and
{\tt NAAMY21} polarized PDFs with statistical uncertainty are also shown for comparison. }
\label{Lorentzforce}
\end{table*}

\section{Conclusions} \label{concl}

The following conclusions can be drawn from the present study.
In this article, we have conducted an investigation into the impact 
of symmetry breaking in polarized light sea quarks on the extraction of 
polarized PDFs.
We analyze their effects on the polarized structure functions 
and their moments for both 
nucleons and nuclei, providing a detailed discussion. 
Additionally, we explore and present the implications of symmetry 
breaking on the sum rules and the components of the Lorentz color force. 
Furthermore, it is important to note that in our analysis, 
the polarized structure functions of nuclei are computed while 
considering the effective nuclear modifications. This ensures a more 
accurate description of the nuclear effects on the polarized structure functions.

To investigate the impact of symmetry breaking in polarized light sea quarks, 
we compare the polarized PDFs of the {\tt AKS14}, 
{\tt NAAMY21}, {\tt DSSV09}, and {\tt BB10} models with experimental data 
obtained from {\tt HERMES} and {\tt COMPASS} experiments. 
This comparison allows us to assess the agreement between the theoretical 
models considering symmetry breaking and the experimental data.
The primary outcome of this comparison indicates slightly better agreement 
between the {\tt AKS14} model, which incorporates symmetry breaking, 
and the experimental data. This agreement is particularly notable for 
distributions involving strange and sea quarks. It suggests that considering 
symmetry breaking in the modeling of polarized light sea quarks leads to 
improved agreement with experimental observations.

In addition, we conduct a comparison of the polarized structure 
functions of the proton, neutron, and deuteron using calculations 
from the {\tt AKS14}, {\tt NAAMY21}, {\tt DSSV09}, and {\tt BB10} models. 
This comparison is made with experimental data obtained from the {\tt E143} collaboration. 
The results obtained from the comparisons between data and theory 
indicate that when symmetry breaking is considered, the extracted predictions 
exhibit improved agreement with the experimental data, 
taking into account the associated uncertainties.

To investigate the impact of symmetry breaking on the 
nuclei structure function, we calculate the theoretical predictions 
of the polarized structure function $g_1^{^3\textrm{He}}$ using 
the {\tt NAAMY21}, {\tt AKS14}, and {\tt KTA17} models at both 
NLO and NNLO accuracy in perturbative QCD. 

These theoretical predictions are subsequently compared with 
experimental data from the {\tt E142} and {\tt JLAB} experiments. 
The findings of this comparison indicate that the polarized nuclei 
structure function calculated using the {\tt AKS14} polarized PDFs, 
which consider the symmetry breaking of light sea quarks, exhibits 
better agreement with the experimental data compared 
to other models. This agreement is particularly 
notable in the $0.1 \leq x \leq 0.4$ regions. 

The same findings also apply to the cases 
of $g_2^{^3\textrm{He}}$ and $x^2g_1^{^3\textrm{He}}$ structure functions. 
While the extracted polarized structure functions of nuclei from the {\tt AKS14} model 
demonstrate agreement with the experimental data, we should clarify 
that there are currently no available experimental data for $g_1^{^3\textrm{H}}$. 
Furthermore, considering the uncertainties associated 
with the measurements of $g_2^{^3\textrm{He}}$ and $x^2g_1^{^3\textrm{He}}$, 
it does not allow one to strongly prefer the AKS14 model over other 
models based solely on these measurements.

Our calculations demonstrate that the ratio $\eta$ of the Bjorken sum rule and the extracted Lorentz color
force components $F_{B}^{y,n}$ and $F_{E}^{y,n}$ 
from the {\tt AKS14} model are all in good agreement 
with the measured values as well.

In conclusion, the presented results provide evidence 
that a comprehensive understanding of the spin structure 
of nucleons, nuclei, sum rules, and Lorentz color force 
components of polarized structure functions can often be 
achieved when considering the breaking of both flavors SU(2) 
and SU(3) symmetry in a QCD analysis. By incorporating these 
symmetry-breaking effects, a more accurate description of the 
observed data and a better agreement between 
theory and experiment can be achieved.

\section*{Acknowledgments}

The authors acknowledge the financial support from the Iran 
National Science Foundation (INSF) under grant 
number 4013570. 
S. A. T. and H. K. are also thankful to the School of Particles and Accelerators, 
Institute for Research in Fundamental Sciences (IPM).
F. A. acknowledges the Farhangian University for providing support to conduct this research. 
The authors are grateful to Abolfazl Mirjalili for several useful comments. 
They would also like to express their gratitude to the 
anonymous referee for their valuable comments and suggestions, 
which have greatly contributed to enhancing the quality of this paper.



\newpage




\begin{thebibliography}{99}
\bibitem{Amoroso:2022eow}
S.~Amoroso, A.~Apyan, N.~Armesto, R.~D.~Ball, V.~Bertone, C.~Bissolotti, J.~Bluemlein, R.~Boughezal, G.~Bozzi and D.~Britzger, \textit{et al.}
\href{https://arxiv.org/abs/2203.13923}{[arXiv:2203.13923 [hep-ph]]}.



\bibitem{Hobbs:2020vme}
T.~J.~Hobbs, P.~M.~Nadolsky, F.~I.~Olness and B.~T.~Wang,
\href{https://arxiv.org/abs/2001.07862}{[arXiv:2001.07862 [hep-ph]].}


\bibitem{LHeC:2020van}
P.~Agostini \textit{et al.} [LHeC and FCC-he Study Group],
\href{http://dx.doi.org/10.1088/1361-6471/abf3ba}{{\rm J. Phys. G} {\bfseries 48}, no.11, 110501 (2021)}.
J. Phys. G \textbf{48}, no.11, 110501 (2021),





\bibitem{FCC:2018byv}
A.~Abada \textit{et al.} [FCC],
\href{http://dx.doi.org/10.1140/epjc/s10052-019-6904-3}{{\rm Eur. Phys. J. C} {\bfseries 79}, no.6, 474 (2019)}.

















\bibitem{E142:1996thl}
P.~L.~Anthony \textit{et al.} [E142],
\href{http://dx.doi.org/10.1103/PhysRevD.54.6620}{{\rm Phys. Rev. D} {\bfseries 54}, 6620-6650 (1996)}.


\bibitem{E143:1998hbs}
K.~Abe \textit{et al.} [E143],
\href{http://dx.doi.org/10.1103/PhysRevD.58.112003}{{\rm Phys. Rev. D} {\bfseries 58}, 112003 (1998)}.


\bibitem{COMPASS:2015mhb}
C.~Adolph \textit{et al.} [COMPASS],
\href{http://dx.doi.org/10.1016/j.physletb.2015.11.064}{{\rm Phys. Lett. B} {\bfseries 753}, 18-28 (2016)}.

\bibitem{HERMES:2004zsh}
A.~Airapetian \textit{et al.} [HERMES],
\href{http://dx.doi.org/10.1103/PhysRevD.71.012003}{{\rm Phys. Rev. D} {\bfseries 71}, 012003 (2005)}.

\bibitem{JeffersonLabHallA:2003joy}
X.~Zheng \textit{et al.} [Jefferson Lab Hall A],
\href{http://dx.doi.org/10.1103/PhysRevLett.92.012004}{{\rm Phys. Rev. Lett.} {\bfseries 92}, 012004 (2004)}.


\bibitem{JeffersonLabHallA:2016neg}
D.~Flay \textit{et al.} [Jefferson Lab Hall A],
\href{http://dx.doi.org/10.1103/PhysRevD.94.052003}{{\rm Phys. Rev. D} {\bfseries 94}, no.5, 052003 (2016)}.


\bibitem{Ethier:2020way}
J.~J.~Ethier and E.~R.~Nocera,
\href{http://dx.doi.org/10.1146/annurev-nucl-011720-042725}{{\rm Ann. Rev. Nucl. Part. Sci.} {\bfseries 70}, 43-76 (2020)}.

\bibitem{Deur:2018roz}
A.~Deur, S.~J.~Brodsky and G.~F.~De T\'eramond,
\href{https://arxiv.org/abs/1807.05250}{[arXiv:1807.05250 [hep-ph]].}




\bibitem{Zhou:2022wzm}
Y.~Zhou \textit{et al.} [Jefferson Lab Angular Momentum (JAM)],
\href{http://dx.doi.org/10.1103/PhysRevD.105.074022}{{\rm Phys. Rev. D } {\bfseries 105}, no.7, 074022 (2022)}.





\bibitem{Lin:2017snn}
H.~W.~Lin, E.~R.~Nocera, F.~Olness, K.~Orginos, J.~Rojo, A.~Accardi, C.~Alexandrou, A.~Bacchetta, G.~Bozzi and J.~W.~Chen, \textit{et al.}
\href{http://dx.doi.org/10.1016/j.ppnp.2018.01.007}{{\rm Prog. Part. Nucl. Phys.} {\bfseries 100}, 107-160 (2018)}.



\bibitem{Geesaman:2018ixo}
D.~F.~Geesaman and P.~E.~Reimer,
\href{http://dx.doi.org/10.1088/1361-6633/ab05a7}{{\rm Rept. Prog. Phys.} {\bfseries 82}, no.4, 046301 (2019)}.







\bibitem{CiofidegliAtti:1993zs}
C.~Ciofi degli Atti, S.~Scopetta, E.~Pace and G.~Salme,
\href{http://dx.doi.org/10.1103/PhysRevC.48.R968}{{\rm Phys. Rev. C} {\bfseries 48}, R968-R972 (1993)}.


\bibitem{Bissey:2000ed}
F.~R.~P.~Bissey, A.~W.~Thomas and I.~R.~Afnan,
\href{http://dx.doi.org/10.1103/PhysRevC.64.024004}{{\rm Phys. Rev. C} {\bfseries 64}, 024004 (2001)}.

\bibitem{Afnan:2003vh}
I.~R.~Afnan, F.~R.~P.~Bissey, J.~Gomez, A.~T.~Katramatou, S.~Liuti, W.~Melnitchouk, G.~G.~Petratos and A.~W.~Thomas,
\href{http://dx.doi.org/10.1103/PhysRevC.68.035201}{{\rm Phys. Rev. C} {\bfseries 68}, 035201 (2003)}.

\bibitem{AtashbarTehrani:2007odq}
S.~Atashbar Tehrani and A.~N.~Khorramian,
\href{http://dx.doi.org/10.1088/1126-6708/2007/07/048}{{\rm JHEP} {\bfseries 07}, 048 (2007)}.


\bibitem{Khorramian:2010qa}
A.~N.~Khorramian, S.~Atashbar Tehrani, S.~Taheri Monfared, F.~Arbabifar and F.~I.~Olness,
\href{http://dx.doi.org/10.1103/PhysRevD.83.054017}{{\rm Phys. Rev. D} {\bfseries 83}, 054017 (2011)}.


\bibitem{Taghavi-Shahri:2016idw}
F.~Taghavi-Shahri, H.~Khanpour, S.~Atashbar Tehrani and Z.~Alizadeh Yazdi,
\href{http://dx.doi.org/10.1103/PhysRevD.93.114024}{{\rm Phys. Rev. D} {\bfseries 93},  no.11, 114024 (2016)}.


\bibitem{AtashbarTehrani:2013qea}
S.~Atashbar Tehrani, F.~Taghavi-Shahri, A.~Mirjalili and M.~M.~Yazdanpanah,
\href{http://dx.doi.org/10.1103/PhysRevD.87.114012}{{\rm Phys. Rev. D} {\bfseries 87},  no.11, 114012 (2013)}.
[erratum: Phys. Rev. D \textbf{88}, no.3, 039902 (2013)]


\bibitem{Khanpour:2017cha}
H.~Khanpour, S.~T.~Monfared and S.~Atashbar Tehrani,
\href{http://dx.doi.org/10.1103/PhysRevD.95.074006}{{\rm Phys. Rev. D} {\bfseries 95},  no.7, 074006 (2017)}.


\bibitem{Salajegheh:2018hfs}
M.~Salajegheh, S.~M.~Moosavi Nejad, M.~Nejad, H.~Khanpour and S.~Atashbar Tehrani,
\href{http://dx.doi.org/10.1103/PhysRevC.97.055201}{{\rm Phys. Rev. C} {\bfseries 97},  no.5, 055201 (2018)}.


\bibitem{Nematollahi:2021ynm}
H.~Nematollahi, P.~Abolhadi, S.~Atashbar, A.~Mirjalili and M.~M.~Yazdanpanah,
\href{http://dx.doi.org/10.1140/epjc/s10052-020-08810-1}{{\rm Eur. Phys. J. C} {\bfseries 81},  no.1, 18 (2021)}.



\bibitem{Mirjalili:2022cal}
A.~Mirjalili and S.~Tehrani Atashbar,
\href{http://dx.doi.org/10.1103/PhysRevD.105.074023}{{\rm Phys. Rev. D} {\bfseries 105},  no.7, 074023 (2022)}.



\bibitem{Sidorov:2006fi}
A.~V.~Sidorov and D.~B.~Stamenov,
\href{http://dx.doi.org/10.1142/S0217732306021402}{{\rm Mod. Phys. Lett. A} {\bfseries 21},  1991-1998 (2006)}.






\bibitem{Leader:2014uua}
E.~Leader, A.~V.~Sidorov and D.~B.~Stamenov,
\href{http://dx.doi.org/10.1103/PhysRevD.91.054017}{{\rm Phys. Rev. D} {\bfseries 91},  no.5, 054017 (2015)}.



\bibitem{deFlorian:2005mw}
D.~de Florian, G.~A.~Navarro and R.~Sassot,
\href{http://dx.doi.org/10.1103/PhysRevD.71.094018}{{\rm Phys. Rev. D} {\bfseries 71},  094018 (2005)}.





\bibitem{Gluck:2000dy}
M.~Gluck, E.~Reya, M.~Stratmann and W.~Vogelsang,
\href{http://dx.doi.org/10.1103/PhysRevD.63.094005}{{\rm Phys. Rev. D} {\bfseries 63},  094005 (2001)}.





\bibitem{Blumlein:2002qeu}
J.~Blumlein and H.~Bottcher,
\href{http://dx.doi.org/10.1016/S0550-3213(02)00342-5}{{\rm Nucl. Phys. B} {\bfseries 636},  225-263 (2002)}.






\bibitem{deFlorian:2008mr}
D.~de Florian, R.~Sassot, M.~Stratmann and W.~Vogelsang,
\href{http://dx.doi.org/10.1103/PhysRevLett.101.072001}{{\rm Phys. Rev. Lett.} {\bfseries 101},  072001 (2008)}.


\bibitem{Hirai:2008aj}
M.~Hirai \textit{et al.} [Asymmetry Analysis],
\href{http://dx.doi.org/10.1016/j.nuclphysb.2008.12.026}{{\rm Nucl. Phys. B} {\bfseries 813},  106-122 (2009)}.


\bibitem{Blumlein:2010rn}
J.~Blumlein and H.~Bottcher,
\href{http://dx.doi.org/10.1016/j.nuclphysb.2010.08.005}{{\rm Nucl. Phys. B} {\bfseries 841},  205-230 (2010)}.


\bibitem{Leader:2011tm}
E.~Leader, A.~V.~Sidorov and D.~B.~Stamenov,
\href{http://dx.doi.org/10.1103/PhysRevD.84.014002}{{\rm Phys. Rev. D} {\bfseries 84},  014002 (2011)}.


\bibitem{Leader:2010dx}
E.~Leader, A.~V.~Sidorov and D.~B.~Stamenov,
\href{https://arxiv.org/abs/1007.4781}{[arXiv:1007.4781 [hep-ph]]}.


\bibitem{Leader:2009tr}
E.~Leader, A.~V.~Sidorov and D.~B.~Stamenov,
\href{http://dx.doi.org/10.1103/PhysRevD.80.054026}{{\rm Phys. Rev. D} {\bfseries 80},  054026 (2009)}.


\bibitem{Nocera:2014uea}
E.~R.~Nocera,
\href{http://dx.doi.org/10.1016/j.physletb.2015.01.021}{{\rm Phys. Lett. B} {\bfseries 742},  117-125  (2015)}.


\bibitem{Khanpour:2017fey}
H.~Khanpour, S.~T.~Monfared and S.~Atashbar Tehrani,
\href{http://dx.doi.org/10.1103/PhysRevD.96.074037}{{\rm Phys. Rev. D} {\bfseries 96},  no.7, 074037 (2017)}.


\bibitem{Leader:2010rb}
E.~Leader, A.~V.~Sidorov and D.~B.~Stamenov,
\href{http://dx.doi.org/10.1103/PhysRevD.82.114018}{{\rm Phys. Rev. D} {\bfseries 82}, 1140187 (2010)}.


\bibitem{deFlorian:2009vb}
D.~de Florian, R.~Sassot, M.~Stratmann and W.~Vogelsang,
\href{http://dx.doi.org/10.1103/PhysRevD.80.034030}{{\rm Phys. Rev. D} {\bfseries 80}, 034030 (2009)}.


\bibitem{deFlorian:2014yva}
D.~de Florian, R.~Sassot, M.~Stratmann and W.~Vogelsang,
\href{http://dx.doi.org/10.1103/PhysRevLett.113.012001}{{\rm Phys. Rev. Lett. } {\bfseries 113},no.1, 012001 (2014)}.


\bibitem{Ball:2013lla}
R.~D.~Ball \textit{et al.} [NNPDF],
\href{http://dx.doi.org/10.1016/j.nuclphysb.2013.05.007}{{\rm Nucl. Phys. B} {\bfseries 874}, 36-84 (2013)}.


\bibitem{Arbabifar:2013tma}
F.~Arbabifar, A.~N.~Khorramian and M.~Soleymaninia,
\href{http://dx.doi.org/10.1103/PhysRevD.89.034006}{{\rm Phys. Rev. D} {\bfseries 89}, no.3, 034006 (2014)}.





\bibitem{Khorramian:2020gkr}
A.~Khorramian, E.~Leader, D.~B.~Stamenov and A.~Shabanpour,
\href{http://dx.doi.org/10.1103/PhysRevD.103.054003}{{\rm Phys. Rev. D} {\bfseries 103}, no.5, 054003 (2021)}.











\bibitem{Vogt:2004ns}
A.~Vogt,
\href{http://dx.doi.org/10.1016/j.cpc.2005.03.103}{{\rm Comput. Phys. Commun.} {\bfseries 170}, 65-92 (2005)}.

\bibitem{Lampe:1998eu}
B.~Lampe and E.~Reya,
\href{http://dx.doi.org/10.1016/S0370-1573(99)00100-3}{{\rm Phys. Rept.} {\bfseries 332}, 1-163 (2000)}.

\bibitem{Lacombe:1981eg}
M.~Lacombe, B.~Loiseau, R.~Vinh Mau, J.~Cote, P.~Pires and R.~de Tourreil,
\href{http://dx.doi.org/10.1016/0370-2693(81)90659-6}{{\rm Phys.\ Lett.\ B} {\bfseries 101}, 139 (1981)}.



\bibitem{Buck:1979ff}
W.~W.~Buck and F.~Gross,
\href{http://dx.doi.org/10.1103/PhysRevD.20.2361}{{\rm Phys.\ Rev.\ D} {\bfseries 20}, 2361 (1979)}.



\bibitem{Zuilhof:1980ae}
M.~J.~Zuilhof and J.~A.~Tjon,
\href{http://dx.doi.org/10.1103/PhysRevC.22.2369}{{\rm Phys.\ Rev.\ C} {\bfseries 22}, 2369 (1980)}.



\bibitem{Parisi:1978jv}
G.~Parisi and N.~Sourlas,
\href{http://dx.doi.org/10.1016/0550-3213(79)90448-6}{{\rm Nucl. Phys. B} {\bfseries 151}, 421-428 (1979)}.






\bibitem{Hou:2019efy}
T.~J.~Hou, J.~Gao, T.~J.~Hobbs, K.~Xie, S.~Dulat, M.~Guzzi, J.~Huston, P.~Nadolsky, J.~Pumplin and C.~Schmidt, \textit{et al.}
\href{http://dx.doi.org/10.1103/PhysRevD.103.014013}{{\rm Phys. Rev. D} {\bfseries 103},  no.1, 014013 (2021)}.





\bibitem{Kataev:2001kk}
A.~L.~Kataev, G.~Parente and A.~V.~Sidorov,
\href{http://dx.doi.org/10.1134/S1063779607060068}{{\rm Phys.\ Part.\ Nucl.} {\bfseries 34}, 20 (2003)}.
[Fiz.\ Elem.\ Chast.\ Atom.\ Yadra {\bf 34}, 43 (2003)]
[Phys.\ Part.\ Nucl.\  {\bf 38}, no. 6, 827 (2007)].



\bibitem{Kataev:2005ci}
A.~L.~Kataev,
\href{http://dx.doi.org/10.1134/1.2034588}{{\rm JETP Lett.} {\bfseries 81}, 608 (2005)}.
[Pisma Zh.\ Eksp.\ Teor.\ Fiz.\  {\bf 81}, 744 (2005)].


\bibitem{COMPASS:2010hwr}
M.~G.~Alekseev \textit{et al.} [COMPASS],
\href{http://dx.doi.org/10.1016/j.physletb.2010.08.034}{{\rm Phys. Lett. B} {\bfseries 693},  227-235  (2010)}.






\bibitem{Yazdanpanah:2009zz}
M.~M.~Yazdanpanah, A.~Mirjalili, S.~Atashbar Tehrani and F.~Taghavi-Shahri,
\href{http://dx.doi.org/10.1016/j.nuclphysa.2009.10.080}{{\rm Nucl. Phys. A} {\bfseries 831}, 243-262 (2009)}.



\bibitem{Lacombe:1980dr}
M.~Lacombe, B.~Loiseau, J.~M.~Richard, R.~Vinh Mau, J.~Cote, P.~Pires and R.~De Tourreil,
\href{http://dx.doi.org/10.1103/PhysRevC.21.861}{{\rm Phys. Rev. C } {\bfseries 21}, 861-873 (1980)}.


\bibitem{Frankfurt:1996nf}
L.~Frankfurt, V.~Guzey and M.~Strikman,
\href{http://dx.doi.org/10.1016/0370-2693(96)00625-9}{{\rm Phys. Lett. B} {\bfseries 381},  379-384 (1996)}.


\bibitem{Saito:1990aj}
T.~Y.~Saito, Y.~Wu, S.~Ishikawa and T.~Sasakawa,
\href{http://dx.doi.org/10.1016/0370-2693(90)91586-Z}{{\rm Phys. Lett. B} {\bfseries 242},  12-16 (1990)}.


\bibitem{Carlson:1991ju}
J.~Carlson, D.~O.~Riska, R.~Schiavilla and R.~B.~Wiringa,
\href{http://dx.doi.org/10.1103/PhysRevC.44.619}{{\rm Phys. Rev. C} {\bfseries 44}, 619-625 (1991)}.


\bibitem{Boros:2000af}
C.~Boros, V.~A.~Guzey, M.~Strikman and A.~W.~Thomas,
\href{http://dx.doi.org/10.1103/PhysRevD.64.014025}{{\rm Phys. Rev. D} {\bfseries 64}, 014025 (2001)}.


\bibitem{Bissey:2001cw}
F.~R.~P.~Bissey, V.~A.~Guzey, M.~Strikman and A.~W.~Thomas,
\href{http://dx.doi.org/10.1103/PhysRevC.65.064317}{{\rm Phys. Rev. C} {\bfseries 65}, 064317 (2002)}.


\bibitem{Ethier:2013hna}
J.~J.~Ethier and W.~Melnitchouk,
\href{http://dx.doi.org/10.1103/PhysRevC.88.054001}{{\rm Phys. Rev. C} {\bfseries 88}, no.5, 054001 (2013)}.


\bibitem{JeffersonLabHallA:2004tea}
X.~Zheng \textit{et al.} [Jefferson Lab Hall A],
\href{http://dx.doi.org/10.1103/PhysRevC.70.065207}{{\rm Phys. Rev. C} {\bfseries 70}, 065207 (2004)}.


\bibitem{jlab:WilliamandMary}
K. Kramer, Ph.D. thesis, College of William and Mary, 2003.
\bibitem{Wandzura:1977qf}
S.~Wandzura and F.~Wilczek,
\href{http://dx.doi.org/10.1016/0370-2693(77)90700-6}{{\rm Phys. Lett. B} {\bfseries 72},  195-198 (1977)}.


\bibitem{Budick:1991zb}
B.~Budick, J.~S.~Chen and H.~Lin,
\href{http://dx.doi.org/10.1103/PhysRevLett.67.2630}{{\rm Phys. Rev. Lett.} {\bfseries 67},  2630-2633 (1991)}.



\bibitem{Bjorken:1966jh}
J.~D.~Bjorken,
\href{http://dx.doi.org/10.1103/PhysRev.148.1467}{{\rm Phys. Rev.} {\bfseries 148},  1467-1478 (1966)}.



\bibitem{ParticleDataGroup:2016lqr}
C.~Patrignani \textit{et al.} [Particle Data Group],
\href{http://dx.doi.org/10.1088/1674-1137/40/10/100001}{{\rm Chin. Phys. C} {\bfseries 40},  no.10, 100001 (2016)}.



\bibitem{blum:1997}
J.~Blumlein and N.~Kochelev,
\href{http://dx.doi.org/10.1016/S0550-3213(97)00234-4}{{\rm Nucl. Phys. B} {\bfseries 498},  285-309  (1997)}.



\bibitem{E155:2003}
P.~L.~Anthony \textit{et al.} [E155],
\href{http://dx.doi.org/10.1016/S0370-2693(02)03015-0}{{\rm Phys. Lett. B} {\bfseries 553},  18-24 (2003)}.


\bibitem{Burkardt:2008ps}
M.~Burkardt,
\href{https://link.aps.org/doi/10.1103/PhysRevD.88.114502}{{\rm Phys. Rev. D} {\bfseries 88}, 114502 (2013)}.


\bibitem{Kataev:1994gd}
A.~L.~Kataev,
\href{http://dx.doi.org/10.1103/PhysRevD.50.R5469}{{\rm Phys. Rev. D} {\bfseries 50}, R5469-R5472 (1994)}.




\bibitem{Larin:1997qq}
S.~A.~Larin, T.~van Ritbergen and J.~A.~M.~Vermaseren,
\href{http://dx.doi.org/10.1016/S0370-2693(97)00534-0}{{\rm Phys. Lett. B} {\bfseries 404}, 153-160 (1997)}.



\bibitem{Osipenko:2004xg}
M.~Osipenko, W.~Melnitchouk, S.~Simula, P.~E.~Bosted, V.~Burkert, M.~E.~Christy, K.~Griffioen, C.~Keppel and S.~E.~Kuhn,
\href{http://dx.doi.org/10.1016/j.physletb.2005.01.071}{{\rm Phys. Lett. B} {\bfseries 609}, 259-264 (2005)}.



\bibitem{Meziani:2004ne}
Z.~E.~Meziani, W.~Melnitchouk, J.~P.~Chen, S.~Choi, T.~Averett, G.~Cates, C.~W.~de Jager, A.~Deur, H.~Gao and F.~Garibaldi, \textit{et al.}
\href{http://dx.doi.org/10.1016/j.j.physletb.2005.03.046}{{\rm Phys. Lett. B} {\bfseries 613}, 148-153 (2005)}.









\end{thebibliography}
\end{document}